\documentclass[12pt]{article}

\usepackage{amsmath}
\usepackage{amssymb}
\usepackage{enumerate}
\usepackage{indentfirst}
\begin{document}

\title{On the additivity conjecture for channels with arbitrary constraints}
\author{M.E.Shirokov}
\date{}
\maketitle

\section{Introduction}

In \cite{Sh-e-a-q} Shor proved equivalence of several open (sub)additivity
problems related to the Holevo capacity and the entanglement of formation
(EoF). In \cite{H-Sh} we showed equivalence of these to the additivity of
the Holevo capacity for channels with arbitrary linear constraints.

This note is the development of \cite{H-Sh} in the direction of channels
with general constraints. Introducing input constraints provides greater
flexibility in the treatment of the additivity conjecture. On the other
hand, while \cite{Sh-e-a-q}, \cite{H-Sh} deal with the ''global'' additivity
conjecture, i.e. properties valid for all possible channels, in this paper
we make emphasis on results valid for individual channels. The Holevo
capacity of the arbitrarily constrained channel is considered and the
characteristic property of an optimal ensemble for such channel is derived
(proposition 1), generalizing the maximal distance property of Schumacher
and Westmoreland \cite{Sch-West-1} . This property provides a useful
estimate for the Holevo capacity of the constrained channel (proposition 2).
Our main point of interest is the additivity conjecture for two arbitrarily
constrained channels and its relations to Shor's channel extension \cite
{Sh-e-a-q}. An attempt to prove converse of the theorem 1 in \cite{H-Sh}, in
which sufficient condition for ''constrained additivity'' was formulated,
leads to the notion of asymptotic additivity for sequences of channels. This
notion applied to the sequences of Shor's channel extensions makes it
possible to formulate the main result of \cite{H-Sh} in the form of
necessary and sufficient condition. This implies that the additivity
conjecture for two channels with \textit{single} linear constraints is
equivalent to the similar conjecture for two \textit{arbitrarily}
constrained channels and, hence, to an interesting subadditivity property of
the $\chi $-function for the tensor product of these channels.

This subadditivity property of the $\chi$-function seems to be very
appealing. It implies the unconstrained additivity and can be used for
proving the latter in some cases. This property can be established for
several types of channels (proposition 4). The characteristic property of
the optimal ensembles for constrained channels provides necessary and
sufficient condition for additivity of the Holevo capacity for two channels,
constrained by fixing partial states of average state of input ensemble
(theorem 2). This leads to an interesting characterization of the channels
for which subadditivity of the $\chi$-function holds (corollary 5).

Global subadditivity of the $\chi $-function follows obviously from the
strong superadditivity of the entanglement of formation and from the MSW
channel representation \cite{MSW}. But it turns out that the converse is
also true (corollary 6). With the corollary 3 this gives another (in
comparison to \cite{Sh-e-a-q}) way of proving that the global additivity
conjecture for unconstrained channels implies strong superadditivity of the
entanglement of formation.

The arguments from the convex analysis raised in \cite{A&B} provide another
characterization of channels for which subadditivity of the $\chi $-function
holds (theorem 3). This characterization and some modification of Shor's
channel extension provides a simple way of proving that global additivity of
the minimum output entropy for unconstrained channels implies global
subadditivity of the $\chi $-function and strong superadditivity of the
entanglement of formation.

\section{Channels with arbitrary constraints}

Let $\Phi :\mathfrak{S}(\mathcal{H})\mapsto \mathfrak{S}(\mathcal{H}^{\prime
})$ be an arbitrary channel (here $\mathcal{H},\mathcal{H}^{\prime}$ are
finite dimensional Hilbert spaces). Let $\{\rho _{i}\}$ be an arbitrary
ensemble of states in $\mathfrak{S}( \mathcal{H})$ with the probability
distribution $\{\pi _{i}\}$, which will be denoted as $\{\pi _{i}, \rho
_{i}\}$.

The Holevo quantity \cite{H} for this ensemble is defined by
\[
\chi _{\Phi }\left( \{\pi _{i}, \rho _{i}\}\right) =H\left( \sum_{i}\pi
_{i}\Phi \left( \rho _{i}\right) \right) -\sum_{i}\pi _{i}H\left( \Phi
\left( \rho _{i}\right) \right).
\]
For an arbitrary state $\rho$ we denote
\[
\chi _{\Phi }(\rho )= \max\chi_{\Phi}(\{\pi_{i},\rho_{i}\}),\vspace{10pt}
\]
where the maximum is over all ensembles $\{\pi _{i}, \rho _{i}\}$ with $
\sum_{i}\pi _{i}\rho _{i}=\rho $. The function $\chi _{\Phi }(\rho )$
(briefly $\chi$-function) is continuous and concave on $\mathfrak{S}(
\mathcal{H})$. The concavity of $\chi _{\Phi }(\rho )$ easily follows from
the definition. The continuity of $\chi _{\Phi }(\rho )$ can be derived from
the MSW-correspondence \cite{MSW} and the continuity of the entanglement of
formation \cite{N}.

Fixing a closed subset $\mathcal{A}$ of $\mathfrak{S}(\mathcal{H})$ we can
consider the constraints on the input ensemble $\{\pi _{i},\rho _{i}\}$ of
the channel $\Phi $ defined by the requirement $\rho _{\mathrm{av}}\in
\mathcal{A}$, where $\rho _{\mathrm{av}}=\sum_{i}\pi _{i}\rho _{i}$ is the
average state of the ensemble. This type of constraints is a natural
generalization of linear constraints considered in \cite{H-c-w-c}, where the
subset $\mathcal{A}$ is defined by the inequality $\mathrm{Tr}A\rho _{
\mathrm{av}}\leq \alpha $ for some positive operator $A$ and positive number
$\alpha $.

Define the Holevo capacity of the $\mathcal{A}$-constrained channel $\Phi$
by
\begin{equation}
\bar{C}(\Phi ;\mathcal{A})=\max_{\rho\in\mathcal{A} }\chi _{\Phi }(\rho
)=\max_{\sum\pi_{i}\rho_{i}\in\mathcal{A} }\chi _{\Phi }(\{\pi_{i},
\rho_{i}\}).  \label{ccap}
\end{equation}
Note that the unconstrained capacity $\bar{C}(\Phi)=\bar{C}(\Phi ;
\mathfrak{S }(\mathcal{H}))$.

Any ensemble $\{\pi _{i},\rho _{i}\}$ on which the maximum in (\ref{ccap})
is achieved is called an \textit{optimal ensemble} for the $\mathcal{A}$
-constrained channel $\Phi $. In \cite{Sch-West-1} it was shown that an
optimal ensemble $\{\pi _{i},\rho _{i}\}$ (with the average state $\rho _{
\mathrm{av}}$) for the unconstrained channel $\Phi $ is characterized the by
the \textit{maximal distance property}:
\[
S(\Phi (\omega )\Vert \Phi (\rho _{\mathrm{av}}))\leq \chi _{\Phi }(\{\pi
_{i},\rho _{i}\}),\quad \forall \omega \in \mathfrak{S}(\mathcal{H}).
\]
The generalization of the above property for the constrained channels is
given by the following proposition.

\textbf{Proposition 1.} The ensemble $\{\pi_{i}, \rho_{i}\}$ with the
average state $\rho_{\mathrm{av}}\in \mathcal{A}$ is optimal for the $
\mathcal{A}$-constrained channel $\Phi$ with a \textit{convex} set $\mathcal{
\ A}$ if and only if
\[
\sum_{j}\mu_{j}S(\Phi(\omega_{j})\|\Phi(\rho_{\mathrm{av}}))\leq
\chi_{\Phi}(\{\pi_{i}, \rho_{i}\})
\]
for any ensemble $\{\mu_{j}, \omega_{j}\}$ of states in $\mathfrak{S}(
\mathcal{H})$ with the average state $\omega_{\mathrm{av}}\in \mathcal{A}$.

\textit{Proof.} The proof is a simple generalization of the arguments in
\cite{Sch-West-1}. The specific feature of the "constrained" situation
consists in the necessity to consider variations of the initial ensemble by
mixing not only one state as in \cite{Sch-West-1}, but a genuine ensemble.

Let $\{\pi _{i},\rho _{i}\}_{i=1}^{n}$ and $\{\mu _{j},\omega
_{j}\}_{j=1}^{m}$ be two ensembles with average states $\rho _{\mathrm{av}}$
and $\omega _{\mathrm{av}}$, contained in $\mathcal{A}$. Consider the
modification of the first ensemble by adding the second one with the weight
coefficient $\eta $. The modified ensemble
$$
\{ (1-\eta )\pi _{1}\rho _{1},...,(1-\eta )\pi _{n}\rho _{n},\eta \mu
_{1}\omega _{1},...,\eta \mu _{m}\omega _{m}\}
$$
has the average state $\rho _{\mathrm{av}}^{\eta }=(1-\eta )\rho _{\mathrm{
av }}+\eta \omega _{\mathrm{av}}$, contained in $\mathcal{A}$ (by
convexity). Let $\chi _{\Phi }$ and $\chi _{\Phi }^{\eta }$ be the Holevo
quantities for the original and the modified ensembles correspondingly.
Using the relative entropy expression for the Holevo quantity \cite
{Sch-West-2} we have
\begin{equation}
\chi _{\Phi }^{\eta }=(1-\eta )\sum_{i=1}^{n}\pi _{i}S(\Phi (\rho _{i})\Vert
\Phi (\rho _{\mathrm{av}}^{\eta }))+\eta \sum_{j=1}^{m}\mu _{j}S(\Phi
(\omega _{j})\Vert \Phi (\rho _{\mathrm{av}}^{\eta })).  \label{m-chi}
\end{equation}
Applying Donald's identity \cite{Sch-West-1},\cite{Sch-West-2} to the
original ensemble we obtain
\[
\sum_{i=1}^{n}\pi _{i}S(\Phi (\rho _{i})\Vert \Phi (\rho _{\mathrm{av}
}^{\eta }))=\chi _{\Phi }+S(\Phi (\rho _{\mathrm{av}})\Vert \Phi (\rho _{
\mathrm{av}}^{\eta })).
\]
Substitution of the above expression into (\ref{m-chi}) gives
\begin{equation}
\!\!\!\chi _{\Phi }^{\eta }=\chi _{\Phi }+\eta \left[ \sum_{j=1}^{m}\mu
_{j}S(\Phi (\omega _{j})\Vert \Phi (\rho _{\mathrm{av}}^{\eta }))-\chi
_{\Phi }\right] +(1-\eta )S(\Phi (\rho _{\mathrm{av}})\Vert \Phi (\rho _{
\mathrm{av}}^{\eta })).  \label{m-chi-1}
\end{equation}
Applying Donald's identity to the modified ensemble we obtain
\[
\begin{array}{c}
(1-\eta )\sum\limits_{i=1}^{n}\pi _{i}S(\Phi (\rho _{i})\Vert \Phi (\rho _{
\mathrm{av}}))+\eta \sum\limits_{j=1}^{m}\mu _{j}S(\Phi (\omega _{j})\Vert
\Phi (\rho _{\mathrm{av}})) \\
=\chi _{\Phi }^{\eta }+S(\Phi (\rho _{\mathrm{av}}^{\eta })\Vert \Phi (\rho
_{\mathrm{av}}))
\end{array}
\]
and hence
\begin{equation}
\chi _{\Phi }^{\eta }=\chi _{\Phi }+\eta \left[ \sum_{j=1}^{m}\mu _{j}S(\Phi
(\omega _{j})\Vert \Phi (\rho _{\mathrm{av}}))-\chi _{\Phi }\right] -S(\Phi
(\rho _{\mathrm{av}}^{\eta })\Vert \Phi (\rho _{\mathrm{av}})).
\label{m-chi-2}
\end{equation}

Since the relative entropy is nonnegative the expressions (\ref{m-chi-1})
and (\ref{m-chi-2}) imply the following double inequality for the value $
\Delta\chi_{\Phi}=\chi^{\eta}_{\Phi}-\chi_{\Phi}$:
\begin{equation}  \label{d-chi}
\begin{array}{c}
\eta\left[\sum\limits_{j=1}^{m}\mu_{j}S(\Phi(\omega_{j})\|\Phi(\rho^{\eta}_{
\mathrm{av}}))- \chi_{\Phi}\right] \\
\leq \Delta\chi_{\Phi} \leq \\
\eta\left[\sum\limits_{j=1}^{m}\mu_{j}S(\Phi(\omega_{j})\|\Phi(\rho_{\mathrm{
\ av}}))- \chi_{\Phi}\right].
\end{array}
\end{equation}
Now the proof of the proposition is straightforward. If
\[
\sum_{j}\mu_{j}S(\Phi(\omega_{j})\|\Phi(\rho_{\mathrm{av}}))\leq \chi_{\Phi}
\]
for any ensemble $\{\mu_{j}, \omega_{j}\}$ of states in $\mathfrak{S}(
\mathcal{H})$ with the average state $\omega_{\mathrm{av}}\in \mathcal{A}$,
then by (\ref{d-chi}) with $\eta=1$ we have
\[
\chi_{\Phi}(\{\mu_{j}, \omega_{j}\})=\chi^{1}_{\Phi}\leq
\chi_{\Phi}=\chi_{\Phi}(\{\pi_{i}, \rho_{i}\}),
\]
which means optimality of the ensemble $\{\pi_{i}, \rho_{i}\}$.

To prove the converse, suppose $\{\pi_{i}, \rho_{i}\}$ is an optimal
ensemble and there exists an ensemble $\{\mu_{j}, \omega_{j}\}$ such that
\[
\sum_{j}\mu_{j}S(\Phi(\omega_{j})\|\Phi(\rho_{\mathrm{av}}))> \chi_{\Phi}.
\]
By continuity of the relative entropy there is $\eta>0$ such that
\[
\sum_{j}\mu_{j}S(\Phi(\omega_{j})\|\Phi(\rho^{\eta}_{\mathrm{av}}))>
\chi_{\Phi},
\]
because $\rho^{\eta}_{\mathrm{av}}$ tends to $\rho_{\mathrm{av}}$ when $\eta$
tends to zero. By the first inequality in (\ref{d-chi}), the last inequality
means that $\chi^{\eta}_{\Phi}>\chi_{\Phi}$ in contradiction with the
optimality of the ensemble $\{\pi_{i}, \rho_{i}\}$.$\triangle$

The above characteristic of an optimal ensemble provides the following
estimate for the Holevo capacity of the constrained channel.

\textbf{Proposition 2.} Let $\rho _{\mathrm{av}}$ be the average state of
any optimal ensemble for the $\mathcal{A}$-constrained channel $\Phi $ with
a closed convex subset $\mathcal{A}$ of $\mathfrak{S}(\mathcal{H})$. Then
\[
\bar{C}(\Phi ;\mathcal{A})=\chi _{\Phi }(\rho _{\mathrm{av}})\geq \chi
_{\Phi }(\rho )+S(\Phi (\rho )\Vert \Phi (\rho _{\mathrm{av}})),\quad
\forall \rho \in \mathcal{A}.
\]
\textit{Proof.} Let $\{\pi _{i},\rho _{i}\}$ be an arbitrary ensemble such
that $\sum_{i}\pi _{i}\rho _{i}=\rho \in \mathcal{A}$. By proposition 1
\[
\sum_{i}\pi _{i}S(\Phi (\rho _{i})\Vert \Phi (\rho _{\mathrm{av}}))\leq \chi
_{\Phi }(\rho _{\mathrm{av}}).
\]
This inequality and Donald's identity
\[
\sum_{i}\pi _{i}S(\Phi (\rho _{i})\Vert \Phi (\rho _{\mathrm{av}}))=\chi
_{\Phi }(\{\pi _{i},\rho _{i}\})+S(\Phi (\rho )\Vert \Phi (\rho _{\mathrm{av}
})).
\]
complete the proof.$\triangle $

\section{Shor's channel extension and constrained channels}

Let $\Psi :\mathfrak{S}(\mathcal{K})\mapsto \mathfrak{S}(\mathcal{K}^{\prime
})$ be another channel with the constraint, defined by a closed subset $
\mathcal{B}$ of $\mathfrak{S}(\mathcal{K})$. For the channel $\Phi \otimes
\Psi $ it is natural to consider the constraint defined by the requirements $
\sigma _{\mathrm{av}}^{\Phi }=\mathrm{Tr}_{\mathcal{K}}\sigma _{\mathrm{av}
}\in \mathcal{A}$ and \break $\sigma _{\mathrm{av}}^{\Psi }=\mathrm{Tr}_{
\mathcal{H}}\sigma _{\mathrm{av}}\in \mathcal{B}$ on the average state $
\sigma _{\mathrm{av}}$ of the input ensemble $\{\pi _{i},\sigma _{i}\}$. The
closed subset of $\mathfrak{S}(\mathcal{H}\otimes \mathcal{K})$ defined by
the above requirements will be denoted as $\mathcal{A}\otimes \mathcal{B}$.

We conjecture the following additivity property for constrained channels
\begin{equation}
\bar{C}\left( \Phi \otimes \Psi ;\mathcal{A}\otimes \mathcal{B}\right)
\stackrel{?}{=}\bar{C}(\Phi ;\mathcal{A})+\bar{C}(\Psi ;\mathcal{B}).
\label{addit}
\end{equation}
Note that additivity of the Holevo capacity is a partcular case of (\ref
{addit}) with $\mathcal{A}=\mathfrak{S}(\mathcal{H})$ and $\mathcal{B}=
\mathfrak{S}(\mathcal{K})$.

In \cite{H-Sh} the role of Shor's channel extension \cite{Sh-e-a-q} in
connection with the constrained additivity problem was demonstrated. For
reader's convenience we give here a brief description of this notion.

Let $E$ be an operator in $\mathfrak{B}(\mathcal{H}),0\leq E\leq I_{\mathcal{
\ H}}$ (the identity operator in the space $\mathcal{H}$), let $q\in [0;1]$
and $d\in \mathbb{N}=\{ 1,2,\dots \} .$ Consider the channel $\widehat{\Phi }
(E,q,d)$, which maps $\mathfrak{B}(\mathcal{H} )\otimes \mathbf{C}^{d}$ into
$\mathfrak{B}(\mathcal{H}^{\prime })\oplus \mathbf{C}^{d+1}$, where $\mathbf{
C}^{d}$ is the commutative algebra of complex $d$-dimensional vectors
describing a classical system. By using the isomorphism of $
\mathfrak{B}(\mathcal{H})\otimes \mathbf{C}^{d}$ with the direct sum of $d$
copies of $\mathfrak{B}(\mathcal{H})$, any state in $\mathfrak{B}(
\mathcal{H})\otimes \mathbf{C}^{d}$ can be represented as an array $\{\rho
_{j}\}_{j=1}^{d}$ of positive operators in $\mathfrak{B}( \mathcal{H})$ such
that $\mathrm{Tr}\sum_{j=1}^{d}\rho _{j}=1$. The action of the channel $
\widehat{\Phi }(E,q,d)$ on the state $\widehat{\rho }=\{\rho _{j}\}_{j=1}^{d}
$ with $\rho =\sum_{j=1}^{d}\rho _{j}$ is defined by
\[
\widehat{\Phi }(E,q,d)(\widehat{\rho })=(1-q)\Phi _{0}(\widehat{\rho }
)\oplus q\Phi _{1}(\widehat{\rho }),
\]
where $\Phi _{0}(\widehat{\rho })\!=\Phi (\rho )\in \mathfrak{S}(\mathcal{H}
^{\prime })$ and $\Phi _{1}(\widehat{\rho })=[$\textrm{Tr}$\rho \bar{E},$
\textrm{Tr}$\rho _{1}E,...,$\textrm{Tr}$\rho _{d}E]\in \mathbf{C}^{d+1}\!$
(throughout this paper we use the notation $\bar{A}=I-A$ for operators).
Note that $\Phi _{0}$ and $\Phi _{1}$ are channels from $\mathfrak{B}(
\mathcal{H})\otimes \mathbf{C}^{d}$ to $\mathfrak{B}(\mathcal{H}^{\prime })$
and to $\mathbf{C}^{d+1}$ \break correspondingly. The state space of the
channel $\widehat{\Phi }(E,q,d)$ will be denoted as $\mathfrak{S}_{\widehat{
\Phi}}$.

\pagebreak

We will need some generalization of the proposition 1 in \cite{H-Sh}.

\textbf{Proposition 3.} Let $\Psi :\mathfrak{S}(\mathcal{K})\mapsto
\mathfrak{S}(\mathcal{K}^{\prime })$ be an arbitrary $\mathcal{B}$
-constrained channel. Consider the channel $\widehat{\Phi }(E,q,d)\otimes
\Psi $. Then
\[
\begin{array}{c}
\!\!\!\left\vert \bar{C}\!\left( \widehat{\Phi }(E,q,d)\otimes \Psi ,
\mathfrak{S}_{\widehat{\Phi }}\otimes \mathcal{B}\right)
-\!\!\!\!\max\limits_{\sigma :\mathrm{Tr}_{\mathcal{H}}\sigma \in \mathcal{B}
}\left[ (1\!-\!q)\chi _{\Phi \otimes \Psi }(\sigma )+q\log d\mathrm{Tr}
(E\otimes I_{\mathcal{K}})\,\sigma \right] \right\vert \\
\\
\leq q(\log \dim \mathcal{\mathcal{K}}^{\prime }+1).
\end{array}
\]
\textit{Proof.} Due to the representation
\begin{equation}
\widehat{\Phi }(E,q,d)\otimes \Psi =\left( (1-q)\Phi _{0}\otimes \Psi
\right) \oplus \left( q\Phi _{1}\otimes \Psi \right) ,  \label{d-s-rep}
\end{equation}
the lemma 1 in \cite{H-Sh} reduces the calculation of the Holevo quantity $
\chi _{\widehat{\Phi }(E,q,d)\otimes \Psi }$ for any ensemble of input
states to the calculation of the Holevo quantities $\chi _{\Phi _{k}\otimes
\Psi },\;k=0,1$ for this ensemble:
\begin{equation}
\chi _{\widehat{\Phi }(E,q,d)\otimes \Psi }=(1-q)\chi _{\Phi _{0}\otimes
\Psi }+q\chi _{\Phi _{1}\otimes \Psi }.  \label{chi-exp}
\end{equation}

Note that any state $\widehat{\sigma}$ in $\mathfrak{B}(\mathcal{H})\otimes
\mathbf{C}^{d}\otimes\mathfrak{B}(\mathcal{K})$ can be represented as an
array $\{\sigma _{j}\}_{j=1}^{d}$ of positive operators in $\mathfrak{B}(
\mathcal{H}\otimes\mathcal{K}) $ such that $\mathrm{Tr}\sum_{j=1}^{d}
\sigma_{j}=1$. Denote by $\delta _{j}(\sigma)$ the array $\hat{\sigma}$ with
the state $\sigma$ in the $j$-th position and with zeroes in other places.

It is known \cite{Sch-West-1} that for any channel there exists a pure state
optimal ensemble. This fact and symmetry arguments imply the existence of an
optimal ensemble for the channel $\widehat{\Phi }(E,q,d)\otimes \Psi $
consisting of the states $\widehat{\sigma }_{i,j}=\delta _{j}(\sigma _{i})$
with the probabilities $\widehat{\mu }_{i,j}=d^{-1}\mu _{i}$, where $\{\mu
_{i},\sigma _{i}\}$ is a particular ensemble of states in $\mathfrak{S}(
\mathcal{H\otimes K})$ (cf. \cite{Sh-e-a-q}). Let $\widehat{\sigma }_{
\mathrm{av}}=\sum_{i,j}\widehat{\mu }_{i,j}\widehat{\sigma }_{i,j}$ and $
\sigma _{\mathrm{av}}=\sum_{i}\mu _{i}\sigma _{i}$ be the average states of
these ensembles. Note that $\widehat{\sigma }_{\mathrm{av}}=[d^{-1}\sigma _{
\mathrm{av}},...,d^{-1}\sigma _{\mathrm{av}}]$.

The action of the channel $\Phi _{0}\otimes \Psi $ on the state $\widehat{
\sigma }=\{\sigma _{i}\}_{i=1}^{d}$ with \break $\sigma =\sum_{i=1}^{d}\sigma _{i}$
is defined by
\[
\Phi _{0}\otimes \Psi (\widehat{\sigma })=\Phi \otimes \Psi (\sigma ).
\]
Hence $\Phi _{0}\otimes \Psi (\widehat{\sigma }_{i,j})=\Phi \otimes \Psi
(\sigma _{i})$ and
\begin{equation}
\chi _{\Phi _{0}\otimes \Psi }(\{\widehat{\mu }_{i,j},\widehat{\sigma }
_{i,j}\})=\chi _{\Phi \otimes \Psi }(\{\mu _{i},\sigma _{i}\}).
\label{chi-11}
\end{equation}

Let us prove that
\begin{equation}
\begin{array}{c}
\chi _{\Phi _{1}\otimes \Psi }(\{\widehat{\mu} _{i,j}, \widehat{\sigma }
_{i,j}\})=\log d\mathrm{Tr}(E\otimes I_{\mathcal{K}})\sigma _{\mathrm{av}
}+f_{\Psi}^{E}(\{\mu _{i},\sigma _{i}\}),
\end{array}
\label{chi-12}
\end{equation}
where $f_{\Psi}^{E}(\{\mu _{i},\sigma _{i}\})\leq \log\dim \mathcal{K}
^{\prime}+1$. It is easy to see that the action of the channel $\Phi
_{1}\otimes \Psi$ on the state $\widehat{\sigma}=\{\sigma_{i}\}_{i=1}^{d}$
with $\sigma=\sum_{i=1}^{d}\sigma_{i}$ is defined by
\[
\Phi _{1}\otimes \Psi(\widehat{\sigma})= [\Psi_{\bar{E}}(\sigma),\Psi_{E}(
\sigma_{1}),...,\Psi_{E}(\sigma_{d})],
\]
where $\Psi_{A}(\cdot)=\mathrm{Tr}_{\mathcal{H}}\left(A\otimes I_{\mathcal{K}
}\cdot(\textup{Id}\otimes\Psi)(\cdot)\right)$ is a completely positive map
from $\mathfrak{S}(\mathcal{H\otimes K})$ into $\mathfrak{B}_{+}(\mathcal{\
K^{\prime}})$ ($A=E,\bar{E}$ and $\textup{Id}$ is an identity map on $
\mathfrak{S}(\mathcal{H})$).

Therefore,
\begin{equation}
H(\Phi _{1}\otimes \Psi (\widehat{\sigma }_{i,j}))=H(\Psi _{\bar{E}}(\sigma
_{i}))+H(\Psi _{E}(\sigma _{i})),  \label{H-Phi-12-S-i}
\end{equation}
and
\[
\begin{array}{c}
\Phi _{1}\otimes \Psi (\widehat{\sigma }_{\mathrm{av}})=\sum\limits_{i,j}
\widehat{\mu }_{i,j}\Phi _{1}\otimes \Psi (\widehat{\sigma }_{i,j}) \\
=[\Psi _{\bar{E}}(\sigma_{\mathrm{av}}), d^{-1}\Psi _{E}(\sigma_{\mathrm{av}
}),...,d^{-1}\Psi _{E}(\sigma_{\mathrm{av}})],
\end{array}
\]
Due to this we can conclude that
\begin{equation}
H(\Phi _{1}\otimes \Psi (\widehat{\sigma }_{\mathrm{av}}))=\log d\mathrm{\ \
\mathrm{Tr}}\Psi _{E}(\sigma _{\mathrm{av}})+H(\Psi _{E}(\sigma _{\mathrm{av
} }))+H(\Psi _{\bar{E}}(\sigma _{\mathrm{av}})).  \label{H-Phi-12-S}
\end{equation}
It is easy to see that
\begin{equation}
\mathrm{Tr}\Psi _{E}(\sigma )=\mathrm{Tr}(E\otimes I_{\mathcal{K}})\,\sigma
\label{tr}
\end{equation}
Using (\ref{H-Phi-12-S-i}), (\ref{H-Phi-12-S}) and (\ref{tr}), we obtain
\begin{equation}
\begin{array}{c}
\chi_{\Phi_{1}\otimes\Psi}(\{\widehat{\mu}_{i,j},\widehat{\sigma}_{i,j}\})
=\log d\,\mathrm{Tr}(E\otimes I_{\mathcal{K}})\sigma _{\mathrm{av}}+ \\
\\
H(\Psi _{E}(\sigma _{\mathrm{av}}))+H(\Psi _{\bar{E}}(\sigma _{\mathrm{av}
})) -\sum\limits_{i}\mu _{i}(H(\Psi _{E}(\sigma _{i}))+H(\Psi _{\bar{E}
}(\sigma _{i}))) \\
\\
=\log d\mathrm{Tr}(E\otimes I_{\mathcal{K}})\sigma _{\mathrm{av}}+\chi
_{\Psi _{E}}(\{\mu _{i},\sigma _{i}\})+\chi _{\Psi _{\bar{E}}}(\{\mu
_{i},\sigma _{i}\}).
\end{array}
\label{chi-12+}
\end{equation}
Using the inequalities $0\leq H(S)\leq \mathrm{Tr}S(\log\dim \mathcal{H}
-\log \mathrm{Tr}S)$ for any positive operator $S\in \mathcal{B}(\mathcal{H})
$ and $h_{2}(x)=x\log x+(1-x)\log(1-x)\leq 1$ it is possible to show that
\begin{equation}  \label{est}
f_{\Psi}^{E}(\{\mu _{i},\sigma_{i}\})=\chi _{\Psi_{E}}(\{\mu _{i},\sigma
_{i}\})+ \chi _{\Psi_{\bar{E}}}(\{\mu _{i},\sigma_{i}\})\leq \log\dim
\mathcal{K}^{\prime}+1.
\end{equation}
Applying this estimation to (\ref{chi-12+}) we obtain (\ref{chi-12}).

The expression (\ref{chi-exp}) with (\ref{chi-11}) and (\ref{chi-12}) imply
\[
\begin{array}{c}
\chi_{\widehat{\Phi}(E,q,d)\otimes\Psi}(\{\widehat{\mu}_{i,j},\widehat{
\sigma }_{i,j}\})\! =\!(1-q)\chi_{\Phi_{0}\otimes\Psi}(\{\widehat{\mu}
_{i,j}, \widehat{\sigma}_{i,j}\})+ q\chi_{\Phi_{1}\otimes\Psi}(\{\widehat{\mu
}_{i,j}, \widehat{\sigma}_{i,j}\}) \\
\\
=(1-q)\chi_{\Phi \otimes \Psi }(\{\mu _{i},\sigma _{i}\})+q\log d\mathrm{Tr}
\sigma _{\mathrm{av}}(E\otimes I_{\mathcal{K}})+q f_{\Psi}^{E}(\{\mu
_{i},\sigma _{i}\}).\vspace{-10pt}
\end{array}
\]
The last equality with (\ref{est}) completes the proof.$\triangle$

When dealing with Shor's channel extension in connection with the
constrained additivity problem it is convenient to slightly change notation.
For nonnegative number $p$ and operator $0\leq A\leq I$ we will denote by $
\widehat{\Phi}_{d}(A,p)$ the Shor's extension $\widehat{\Phi}(A, p/\log d,
d) $ of the channel $\Phi$.

\textbf{Definition 1.} We say that the additivity conjecture holds
asymptotically for the sequence of unconstrained channels $\{\Phi_{n}:
\mathfrak{S}(\mathcal{H}_{n})\mapsto \mathfrak{S}(\mathcal{H}^{\prime}_{n})
\} $ and the channel $\Psi:\mathfrak{S}(\mathcal{K})\mapsto \mathfrak{S}(
\mathcal{K}^{\prime })$ with the constraint defined by the set $\mathcal{B}$
if
\[
\lim_{n\rightarrow+\infty}\bar{C}(\Phi_{n}\otimes\Psi, \mathfrak{S}(
\mathcal{ H}_{n})\otimes\mathcal{B})= \lim_{n\rightarrow+\infty}\bar{C}
(\Phi_{n})+\bar{ C}(\Psi,\mathcal{B}),
\]
assuming that the limits exist and are finite.

Note that if the additivity conjecture holds for the channels $\Phi_{n}$ and
$\Psi$ for all sufficiently large $n$ it obviously holds asymptotically for
the sequence $\{\Phi_{n}\}$ and the channel $\Psi$.

\textbf{Theorem 1.} Let $\Phi :\mathfrak{S}(\mathcal{H})\mapsto
\mathfrak{S}( \mathcal{H}^{\prime })$ and $\Psi :\mathfrak{S}(\mathcal{K})
\mapsto \mathfrak{S}(\mathcal{K}^{\prime })$ be arbitrary channels with the
fixed constraint on the second one defined by the set $\mathcal{B}$. The
following statements are equivalent:

\begin{enumerate}[(i)]

\item  The additivity conjecture is true for the channel $\Phi $ with the
constraint $\mathrm{Tr}A\rho \leq \alpha $ for arbitrary $(A,\alpha )$ and
the $\mathcal{B}$-constrained channel $\Psi $;

\item  The additivity conjecture holds asymptotically for the sequence of
the channels $\{\widehat{\Phi }_{d}(A,p)\}_{d\in \mathbb{N}}$ (without
constraints) and the $\mathcal{B}$-constrained channel $\Psi $ for arbitrary
nonnegative number $p$ and arbitrary operator $0\leq A\leq I$;

\item  The additivity conjecture is true for the $\mathcal{A}$-constrained
channel $\Phi $ with arbitrary closed subset $\mathcal{A}$ in $\mathfrak{S}(
\mathcal{H})$ and the $\mathcal{B}$-constrained channel $\Psi $ .
\end{enumerate}

\textit{Proof.} The implication $(\textup{iii})\Rightarrow(\textup{i})$ is
obvious. So it is sufficient to prove $(\textup{i})\Rightarrow(\textup{ii}
) $ and $(\textup{ii})\Rightarrow(\textup{iii})$.

Note, first of all, that for an operator $0\leq A\leq I$ and a number $p\geq
0$ proposition 3 implies
\begin{equation}  \label{asymp-1}
\lim\limits_{d\rightarrow+\infty}\bar{C}(\Phi_{d}(A,p))= \max_{\rho}\left[
\chi_{\Phi }(\rho)+p\mathrm{Tr}A\rho\right]
\end{equation}
and
\begin{equation}  \label{asymp-2}
\lim\limits_{d\rightarrow+\infty}\bar{C}(\Phi_{d}(A,p)\otimes\Psi)=
\max_{\sigma:\mathrm{Tr}_{\mathcal{H}}\sigma\in \mathcal{B}}\left[
\chi_{\Phi\otimes\Psi}(\sigma)+p\mathrm{Tr}(A\otimes I_{\mathcal{K}
})\,\sigma \right]
\end{equation}
correspondingly.

Begin with $(\textup{i})\Rightarrow(\textup{ii})$. Suppose that there
exist an operator $0\leq A\leq I$ and a number $p\geq 0$ such that the
additivity conjecture is not asymptotically true for the sequence $
\{\Phi_{d}(A,p)\}_{d\in \mathbb{N}}$ and the channel $\Psi$. Due to (\ref
{asymp-1}) and (\ref{asymp-2}) it means that
\begin{equation}  \label{p-e}
\!\!\!\max_{\sigma:\mathrm{Tr}_{\mathcal{H}}\sigma\in \mathcal{B}}\left[
\chi_{\Phi\otimes\Psi}(\sigma)+p\mathrm{Tr}(A\otimes I_{\mathcal{K}
})\,\sigma \right]>\max_{\rho}\left[\chi_{\Phi }(\rho)+p\mathrm{Tr}
A\rho\right]+\bar{C} (\Psi,\mathcal{B}).
\end{equation}

Let $\sigma_{*}$ be a maximum point in the left side of the above inequality
and $\alpha=\mathrm{Tr}(A\otimes I_{\mathcal{K}})\,\sigma_{*}$. By the
statement $(\textup{i})$ the additivity conjecture is true for the channel $
\Phi$ with the constraint $\mathrm{Tr}\bar{A}\rho\leq 1-\alpha$ and the $
\mathcal{B}$-constrained channel $\Psi$. So there exist such states $
\sigma^{\Phi}\!$ and $\sigma^{\Psi}\!\in\mathcal{B}$ that $\mathrm{Tr}
A\sigma^{\Phi}\!\geq\alpha$ and $\chi_{\Phi}(\sigma^{\Phi})+\chi_{\Psi}(
\sigma^{\Psi})\geq \chi_{\Phi\otimes\Psi}(\sigma_{*})$ . Hence
\[
\begin{array}{c}
\max\limits_{\sigma}\left[\chi_{\Phi\otimes\Psi}(\sigma)+p\mathrm{Tr}
(A\otimes I_{\mathcal{K}})\,\sigma\right]=\chi_{\Phi\otimes\Psi}(
\sigma_{*})+p\mathrm{Tr}(A\otimes I_{\mathcal{K}})\,\sigma_{*} \\
\leq \chi_{\Phi}(\sigma^{\Phi})+\chi_{\Psi}(\sigma^{\Psi})+ p\mathrm{Tr}
A\sigma^{\Phi}\leq\max\limits_{\rho}\left[\chi_{\Phi }(\rho)+p\mathrm{Tr}
A\rho\right]+\bar{C}(\Psi; \mathcal{B})
\end{array}
\]
in contradiction with (\ref{p-e}).

The proof of $(\textup{ii})\Rightarrow (\textup{iii})$ consists of two
steps. First we will prove the statement ($\textup{iii}$) with the set $
\mathcal{A}$ defined by the system of $n$ inequalities \break $\mathrm{Tr}A_{k}\rho
\leq \alpha _{k}$, $k=\overline{1,n}$ with arbitrary arrays of operators $
\{A_{k}\}_{k=1}^{n}$ and numbers $\{\alpha _{k}\}_{k=1}^{n}$ such that $
0\leq A_{k}\leq I$ and $0\leq \alpha _{k}\leq 1$ for $k=\overline{1,n}$.
Then we will pass to an arbitrary set $\mathcal{A}$.

Let $\mathcal{A}$ be a set of the above special type. Suppose that
the interior of the set $\mathcal{A}$ is nonempty. It implies that
there exist such a state $\rho$ that \break
$\mathrm{Tr}A_{k}\rho<\alpha _{k},k=\overline{1,n}$. It is
sufficient to show that
\begin{equation}
\bar{C}\left( \Phi\otimes
\Psi;\mathcal{A}\otimes\mathcal{B}\right)\leq\bar{C}(\Phi;
\mathcal{A})+\bar{C}(\Psi;\mathcal{B}).  \label{add-c-m}
\end{equation}
Suppose, $">"$ takes place in (\ref{add-c-m}). Then, there exists
an ensemble $\{\mu _{i},\sigma _{i}\}$ in
$\mathfrak{S}(\mathcal{H}\otimes \mathcal{K})$ with the average
state $\sigma _{\mathrm{av}},$ such that $ \mathrm{Tr}A_{k}\sigma
_{\mathrm{av}}^{\Phi}\leq \alpha_{k},k=\overline{1,n},$ $\sigma
_{\mathrm{av}}^{\Psi}\in\mathcal{B}$ and
\begin{equation}
\chi _{\Phi \otimes \Psi }(\{\mu _{i},\sigma_{i}\})>
\bar{C}(\Phi;\mathcal{A})+\bar{C}(\Psi;\mathcal{B}).
\label{c-c-add-c}
\end{equation}

Let $\rho _{\mathrm{av}}$ be the average of the optimal ensemble
for the $\mathcal{A}$-constrained channel $\Phi$  so that
$\bar{C}(\Phi; \mathcal{A})=\chi _{\Phi }(\rho _{\mathrm{av}}).$

Note, that the state $\rho _{\mathrm{av}}$ is the point of maximum of the
concave function $\chi _{\Phi }(\rho )$ with the constraints $\mathrm{Tr}
A_{k}\rho\leq \alpha _{k},k=\overline{1,n}$. By the Kuhn-Tucker theorem \cite
{JT}\footnote{
We use the strong version of this theorem with the Slater condition, which
follows from the assumption that the interior of the set $\mathcal{A}$ is
nonempty.} there exists a set of nonnegative numbers $\{p_{k}\}_{k=1}^{n}$,
such that $\rho _{\mathrm{av}}$ is the point of the global maximum of the
function $\chi _{\Phi }(\rho )-\sum_{k=1}^{n}p_{k}\mathrm{Tr}A_{k}\rho$ and
the following conditions hold
\begin{equation}
p_{k}(\mathrm{Tr}A_{k}\rho _{\mathrm{av}}-\alpha _{k})=0;\quad k=\overline{
1,n }.  \label{c-a-f}
\end{equation}
It is clear that $\rho _{\mathrm{av}}$ is also the point of the global
maximum of the concave function $\chi _{\Phi }(\rho )+\sum_{k=1}^{n}p_{k}
\mathrm{Tr}\bar{A}_{k}\rho$, so that
\begin{equation}
\chi _{\Phi }(\rho )+\sum_{k=1}^{n}p_{k}\mathrm{Tr} \bar{A}_{k}\rho\leq \chi
_{\Phi }(\rho _{\mathrm{av}})+\sum_{k=1}^{n}p_{k}\mathrm{Tr}\bar{A}_{k}\rho
_{\mathrm{av }},\quad \forall \rho \in \mathfrak{S}(\mathcal{H}).
\label{chi-ine}
\end{equation}

Let $p=\|\sum_{i=1}^{n}p_{i}A_{i}\|$. The case $p=0$ means that $\rho _{
\mathrm{av}}$ is the point of the global maximum of $\chi _{\Phi }(\rho )$.
So in this case the channel $\Phi$ is in fact unconstrained. But the
additivity for the unconstrained channel $\Phi$ and the $\mathcal{B}$
-constrained channel $\Psi$ easily follows from the asymptotic additivity
for the sequence $\widehat{\Phi}_{d}(A, 0)\;$ (with arbitrary operator $A$)
and the $\mathcal{B}$-constrained channel $\Psi$, because each channel in
this sequence is equivalent to $\Phi$.

In the case $p>0$ let $A=p^{-1}\sum_{i=1}^{n}p_{i}A_{i}$. Note that $0\leq
A\leq I$. Consider the sequence $\widehat{\Phi}_{d}(A, p)$ of Shor's
extensions of the channel $\Phi$. Assumed asymptotic additivity together
with (\ref{asymp-1}) and (\ref{asymp-2}) implies
\begin{equation}  \label{asymp-e}
\max_{\sigma}\left[\chi_{\Phi\otimes\Psi}(\sigma)+p\mathrm{Tr}(A\otimes I_{
\mathcal{K}})\,\sigma\right]=\max_{\rho}\left[\chi_{\Phi }(\rho)+p\mathrm{Tr}
A\rho\right]+\bar{C}(\Psi; \mathcal{B}).
\end{equation}
Due to (\ref{c-a-f}) and (\ref{chi-ine}) we have
\begin{equation}  \label{asymp-e-1}
\begin{array}{c}
\max\limits_{\rho}\left[\chi_{\Phi}(\rho)+p\mathrm{Tr}A\rho\right]=
\max\limits_{\rho}\left[\chi_{\Phi }(\rho)+\sum\limits_{k=1}^{n}p_{k}\mathrm{
\ Tr}\bar{A}_{k}\rho\right] \\
= \chi_{\Phi}(\rho_{\mathrm{av}})+ \sum\limits_{k=1}^{n}p_{k}\mathrm{Tr}\bar{
A}_{k}\rho _{\mathrm{av}}= \bar{C}(\Phi; \mathcal{A})+\sum
\limits_{k=1}^{n}p_{k}(1-\alpha_{k}).
\end{array}
\end{equation}
Noting that
\[
\mathrm{Tr}(A_{k}\otimes I_{\mathcal{K}})\,\sigma _{\mathrm{av}}=\mathrm{Tr}
A_{k}\sigma_{\mathrm{av}}^{\Phi}\leq \alpha _{k},\quad \;k=\overline{1,n},
\]
we have by (\ref{c-c-add-c})
\[
\begin{array}{c}
\max\limits_{\sigma}\left[\chi_{\Phi\otimes\Psi}(\sigma)+p\mathrm{Tr}
(A\otimes I_{\mathcal{K}})\,\sigma\right]\geq\chi_{\Phi\otimes\Psi}(\sigma_
\mathrm{av})+ p\mathrm{Tr}(A\otimes I_{\mathcal{K}})\,\sigma_\mathrm{av} \\
\!\!=\chi_{\Phi\otimes\Psi}(\sigma_\mathrm{av})+
\sum\limits_{k=1}^{n}p_{k}\, \mathrm{Tr}(\bar{ A}_{k}\otimes I_{\mathcal{K}
})\sigma_\mathrm{av}>\bar{C} (\Phi; \mathcal{A})+\bar{C}(\Psi; \mathcal{B}
)+\sum\limits_{k=1}^{n}p_{k}(1- \alpha_{k}).
\end{array}
\]
The contradiction of the last inequality with (\ref{asymp-e}) and (\ref
{asymp-e-1}) completes the first step of the proof of $(\textup{ii}
)\Rightarrow(\textup{iii})$ under the condition that the interior of the
set $\mathcal{A}$ is nonempty.

The case where the set $\mathcal{A}$ has no inner point may be reduced to
the previous one with the help of the following lemma.

\textbf{Lemma 1.} Let $\{\mathcal{A}_{n}\}_{n\in \mathbb{N}}$ be a
decreasing sequence of closed subsets of $\mathfrak{S}(\mathcal{H})$ and $
\mathcal{A} =\bigcap_{n\in \mathbb{N}}\mathcal{A}_{n}$. If the additivity
conjecture is true for the $\mathcal{A}_{n}$-constrained channel $\Phi $ and
the $\mathcal{\ B}$-constrained channel $\Psi $ for all $n\in \mathbb{N}$
then this conjecture is true for the $\mathcal{A}$-constrained channel $\Phi
$ and the $\mathcal{B}$-constrained channel $\Psi $ as well.

\textit{Proof.} It is sufficient to prove that
\[
\bar{C}(\Phi,\mathcal{A})=\lim_{n\rightarrow+\infty}\bar{C}(\Phi,\mathcal{A}
_{n}),\;\;\; \bar{C}(\Phi\otimes\Psi,\mathcal{A}\otimes\mathcal{B})=
\lim_{n\rightarrow+\infty}\bar{C}(\Phi\otimes\Psi,\mathcal{A}_{n}\otimes
\mathcal{B}).
\]
Due to $\mathcal{A}\otimes\mathcal{B}=\bigcap_{n\in \mathbb{N}}\mathcal{A}
_{n}\otimes\mathcal{B}$ the second equality is equivalent to the first.

The nonnegative sequence $\bar{C}(\Phi ,\mathcal{A}_{n})$ is decreasing, so
the first of the above limits exists and $"\leq "$ is obvious. For each $
n\in \mathbb{N}\;$ let $\rho _{n}$ be the maximum point of the function $
\chi _{\Phi }(\rho )$ with the constraint $\rho \in \mathcal{A}_{n}$. By
compactness argument we can assume that there exists $\lim_{n\rightarrow
+\infty }\rho _{n}=\rho _{\ast }\!\in \mathcal{A}$. The continuity property
of the function $\chi _{\Phi }(\rho )$ gives $\lim_{n\rightarrow +\infty
}\chi _{\Phi }(\rho _{n})=\chi _{\Phi }(\rho _{\ast })$, which proves the $
"="$.$\triangle $

Consider the sequence $\{\mathcal{A}_{m}\}$ of the subsets in $\mathfrak{S}(
\mathcal{H})$, in which subset $\mathcal{A}_{m}$ is defined by the system of
inequalities $\mathrm{Tr}A_{k}\rho\leq \alpha _{k}+1/m,\;k=\overline{1,n}$.
The interior of each $\mathcal{A}_{m}$ is nonempty. By the previous
consideration the additivity conjecture is true for the $\mathcal{A}_{m}$
-constrained channel $\Phi$ and the $\mathcal{B}$-constrained channel $\Psi$
. The lemma 1 gives desired additivity for the $\mathcal{A}$-constrained
channel $\Phi$ and the $\mathcal{B}$-constrained channel $\Psi$.

Now the proof of the additivity for arbitrary set $\mathcal{A}$ is very
simple. Note that any state $\rho $ in $\mathfrak{S}(\mathcal{H})$ may be
considered as a set $\mathcal{A}_{\rho }=\{\rho \}$, defined by a finite
number of linear inequalities. The result of the first step implies
\begin{equation}
\bar{C}\left( \Phi \otimes \Psi ;\mathcal{A}_{\rho }\otimes \mathcal{B}
\right) =\bar{C}(\Phi ;\mathcal{A}_{\rho })+\bar{C}(\Psi ;\mathcal{B})=\chi
_{\Phi }(\rho )+\bar{C}(\Psi ;\mathcal{B}).  \label{add-r}
\end{equation}
Suppose that the additivity conjecture is not true for the $\mathcal{A}$
-constrained channel $\Phi $ and the $\mathcal{B}$-constrained channel $\Psi
$. Then there exists such a state $\sigma \in \mathfrak{S}(\mathcal{H}
\otimes \mathcal{K})$ that $\sigma ^{\Phi }\in \mathcal{A},\;\sigma ^{\Psi
}\in \mathcal{B}$ and
\[
\chi _{\Phi \otimes \Psi }(\sigma )>\bar{C}(\Phi ;\mathcal{A})+\bar{C}(\Psi
; \mathcal{B})\geq \chi _{\Phi }(\sigma ^{\Phi })+\bar{C}(\Psi ;\mathcal{B}).
\]
This inequality contradicts to (\ref{add-r}) with $\rho =\sigma ^{\Phi }$.
The proof of $(\textup{ii})\Rightarrow (\textup{iii})$ is complete. $
\triangle $

The above theorem implies two sorts of results. The equivalence \break $(\textup{
ii })\Leftrightarrow (\textup{iii})$ implies the corollaries 1-3. The
equivalence $(\textup{i})\Leftrightarrow (\textup{iii})$ gives the
corollary 4.

For the verification of the constrained additivity the following sufficient
condition will be convenient (see the proof of the proposition 4(B)).

\textbf{Corollary 1.} If the additivity conjecture holds true for the
unconstrained channels $\widehat{\Phi }_{d}(A,p)$ with arbitrary pair $(A,p)$
and the $\mathcal{B}$-constrained channel $\Psi $ for all sufficiently large
$d$, then the additivity conjecture is true for the $\mathcal{A}$
-constrained channel $\Phi$ and the $\mathcal{B}$-constrained channel $\Psi$
with arbitrary $\mathcal{A}\subset\mathfrak{S}(\mathcal{H})$.

\textbf{Corollary 2.} The additivity for the Shor's channel extensions $
\widehat{\Phi}_{d}(A,p)$ and $\widehat{\Psi}_{e}(B,r)$ with arbitrary pairs $
(A,p)$ and $(B,r)$ for all sufficiently large $d$ and $e$ implies the
additivity for the $\mathcal{A}$-constrained channel $\Phi$ and the \break $
\mathcal{B}$-constrained channel $\Psi$ with arbitrary $\mathcal{A}\subset
\mathfrak{S}(\mathcal{H})$ and $\mathcal{B}\subset\mathfrak{S}(\mathcal{K})$.

\textit{Proof. }This is obtained by double application of the corollary 1. $
\triangle $

\textbf{Corollary 3.} If the additivity conjecture for \textit{any} two
unconstrained channels holds true then it holds for any two channels with
arbitrary constraints.

\textit{Proof.} This follows from corollary 2. $\triangle $

\textbf{Corollary 4.} If the additivity conjecture holds true for the
channels $\Phi$ and $\Psi$ with the \textit{single} linear constraints $
\mathrm{Tr}\rho A \leq \alpha$ and $\mathrm{Tr}\varrho B\leq \beta$\textrm{\
}correspondingly then the additivity conjecture holds true for the $\mathcal{
\ A}$-constrained channel $\Phi$ and the $\mathcal{B}$-constrained channel $
\Psi$ with \textit{arbitrary} $\mathcal{A}\subset\mathfrak{S}(\mathcal{H})$
and $\mathcal{B}\subset\mathfrak{S}(\mathcal{K})$. \hspace{50pt}

\textit{Proof.} Double application of the equivalence $(\textup{i}
)\Leftrightarrow (\textup{iii})$ in the above theorem. $\triangle $

\pagebreak

\section{Subadditivity property of the $\chi$-function}

The additivity of the Holevo capacity for the channels $\Phi $ and $\Psi $
with arbitrary constraints implies the following subadditivity property of
the \break $\chi $-function:
\begin{equation}
\chi _{\Phi \otimes \Psi }(\sigma )\leq \chi _{\Phi }(\sigma ^{\Phi })+\chi
_{\Psi }(\sigma ^{\Psi }),\quad \forall \sigma \in \mathfrak{S}(\mathcal{H}
\otimes \mathcal{K}).  \label{sub-add}
\end{equation}
To see this it is sufficient to take $\mathcal{A}=\{\sigma ^{\Phi }\}$, $
\mathcal{B}=\{\sigma ^{\Psi }\}$ and note that
\[
\begin{array}{c}
\bar{C}(\Phi \otimes \Psi ;\{\sigma ^{\Phi }\}\otimes \{\sigma ^{\Psi
}\})\geq \chi _{\Phi \otimes \Psi }(\sigma ), \\
\quad \\
\bar{C}(\Phi ;\{\sigma ^{\Phi }\})=\chi _{\Phi }(\sigma ^{\Phi }),\quad \bar{
C}(\Psi ;\{\sigma ^{\Psi }\})=\chi _{\Psi }(\sigma ^{\Psi }).
\end{array}
\]
The subadditivity of the function $\chi _{\Phi \otimes \Psi }(\sigma )$
implies its additivity:
\[
\chi _{\Phi \otimes \Psi }(\sigma ^{\Phi }\otimes \sigma ^{\Psi })=\chi
_{\Phi }(\sigma ^{\Phi })+\chi _{\Psi }(\sigma ^{\Psi }),\quad \forall
\sigma ^{\Phi }\in \mathfrak{S}(\mathcal{H}),\;\forall \sigma ^{\Psi }\in
\mathfrak{S}(\mathcal{K}),
\]
which is not obvious as well.

Subadditivity of the function $\chi_{\Phi\otimes\Psi}(\sigma)$ obviously
implies the additivity of the Holevo capacity for these channels. On the
other hand, it is equivalent to the additivity conjecture for the channels $
\Phi$ and $\Psi$ with arbitrary constraints, and, hence, by corollary 4,
with arbitrary single linear constraints. Corollary 3 shows that global
subadditivity of the $\chi$-function is equivalent to the global additivity
of the Holevo capacity for unconstrained channels. However, it is not clear
whether (\ref{sub-add}) is implied by the additivity of the Holevo capacity
for the channels $\Phi$ and $\Psi$.

There are several cases, where (\ref{sub-add}) can be indeed established.

\textbf{Proposition 4.} Let $\Psi$ be arbitrary channel. The inequality (\ref
{sub-add}) holds in each of the following cases:

\begin{description}
\item[(A)]  $\Phi $ is an entanglement breaking channel;

\item[(B)]  $\Phi $ is a noiseless channel;

\item[(C)]  $\Phi $ is a direct sum mixture of a noiseless channel $\textup{
Id}$ and a channel $\Phi _{0}$ such that the function $\chi _{\Phi
_{0}\otimes \Psi }$ is subadditive.
\end{description}

\textit{Proof.} (A) Shor in \cite{Sh-e-b-c} proved the additivity conjecture
for two unconstrained channel if one of them is an entanglement breaking.
But in the proof of this theorem the subadditivity property of the $\chi $
-function was in fact established. It is interesting that in this case we
can directly deduce the subadditivity of the $\chi $-function from
unconstrained additivity with the help of corollary 2. One should only
verify that entanglement breaking property of any channel implies similar
property of Shor's extension for this channel.

(B) The proof of this statement consists of two steps. First we will prove
the additivity conjecture for two channels if one of them is noiseless and $
\{\rho \}$-constrained while the other channel is arbitrary and
unconstrained. Then we will apply the Shor's channel extension to pass to
the $\{\varrho \}$-constrained second channel.

The proof of the first step is the modification of the proof in \cite{H-QI}
of the "unconstrained" additivity for two channels if one of them is
noiseless, which is based on the Groenevold-Lindblad-Ozawa inequality \cite
{O}
\begin{equation}  \label{GLO}
H(\sigma)\leq\sum_{j}p_{j}H(\sigma_{j}),
\end{equation}
where $\sigma$ is a state of a quantum system before measurement, $
\sigma_{j} $ is the state of this system after measurement with yield $j$
and $p_{j}$ is the probability of this yield.

Let $\Phi=\textup{Id}$ and $\rho$ be an arbitrary state in $\mathfrak{S}(
\mathcal{H})$. We want to prove that
\begin{equation}  \label{s-c-add}
\bar{C}(\textup{Id}\otimes\Psi,\{\rho\}\otimes\mathfrak{S}(\mathcal{K}))=
\bar{C}(\textup{Id}, \{\rho\})+\bar{C}(\Psi,\mathfrak{S}(\mathcal{K}))=
H(\rho)+\bar{C}(\Psi)
\end{equation}

Let $\{\mu _{i},\sigma _{i}\}$ be an ensemble of states in $\mathfrak{S}(
\mathcal{H\otimes K})$ with $\sum_{i}\mu _{i}\sigma _{i}^{\Phi }=\rho $. By
subadditivity of quantum entropy we obtain
\begin{equation}  \label{a}
\begin{array}{c}
\chi _{\textup{Id}\otimes \Psi }(\{\mu _{i},\sigma _{i}\})=H(\textup{Id}
\otimes \Psi (\sum\limits_{i}\mu _{i}\sigma _{i}))-\sum\limits_{i}\mu _{i}H(
\textup{Id}\otimes \Psi (\sigma _{i})) \\
\\
\leq H(\rho )+H(\Psi (\sum\limits_{i}\mu _{i}\sigma _{i}^{\Psi
}))-\sum\limits_{i}\mu _{i}H(\textup{Id}\otimes \Psi (\sigma _{i})).
\end{array}
\end{equation}
Consider the measurement, defined by the observable $\{|e_{j}\rangle \langle
e_{j}|\otimes I_{\mathcal{K}}\}$, where $\{|e_{j}\rangle \}$ is an
orthonormal basis in $\mathcal{H}$. By (\ref{GLO}) we obtain
\[
H(\textup{Id}\otimes \Psi (\sigma _{i}))\leq \sum_{j}p_{ij}H(\Psi (\sigma
_{ij}^{\Psi })),\quad \forall i,
\]
where $p_{ij}=\langle e_{j}|\sigma _{i}|e_{j}\rangle $ and $\sigma
_{ij}=p_{ij}^{-1}|e_{j}\rangle \langle e_{j}|\otimes I_{\mathcal{K}}\cdot
\sigma _{i}\cdot |e_{j}\rangle \langle e_{j}|\otimes I_{\mathcal{K}}$. Note
that $\sum_{j}p_{ij}\sigma _{ij}^{\Psi }=\sigma _{i}^{\Psi }$. This and
previous inequality show that two last terms in (\ref{a}) do not exceed $
\chi _{\Psi }(\{\mu _{i}p_{ij},\sigma _{ij}^{\Psi }\})$ and, hence, $\bar{C}
(\Psi )$. With this observation (\ref{a}) implies (\ref{s-c-add}) and,
hence, the first step of the proof is complete.

It follows that the additivity conjecture holds for the (unconstrained)
Shor's extension of the channel $\Psi$ (with arbitrary parameters) and the $
\{\rho\}$-constrained channel $\Phi=\textup{Id}$. The corollary 1 gives the
desired subadditivity property of the function $\chi_{\textup{Id}
\otimes\Psi}$.

(C) Let $\Phi _{q}=q\textup{Id}\oplus (1-q)\Phi _{0}$. For an arbitrary
channel $\Psi $ we have \break $\Phi _{q}\otimes \Psi =q(\textup{Id}\otimes
\Psi )\oplus (1-q)(\Phi _{0}\otimes \Psi )$. By lemma 1 in \cite{H-Sh} with
the subadditivity of the functions $\chi _{\textup{Id}\otimes \Psi }$ and $
\chi _{\Phi _{0}\otimes \Psi }$ we obtain
\[
\begin{array}{c}
\chi _{\Phi _{q}\otimes \Psi }(\sigma )\leq q\chi _{\textup{Id}\otimes \Psi
}(\sigma )+(1-q)\chi _{\Phi _{0}\otimes \Psi }(\sigma ) \\
\\
\leq q\chi _{\textup{Id}}(\sigma ^{\Phi })+q\chi _{\Psi }(\sigma ^{\Psi
})+(1-q)\chi _{\Phi _{0}}(\sigma ^{\Phi })+(1-q)\chi _{\Psi }(\sigma ^{\Psi
}) \\
\\
=qH(\sigma ^{\Phi })+(1-q)\chi _{\Phi _{0}}(\sigma ^{\Phi })+\chi _{\Psi
}(\sigma ^{\Psi })=\chi _{\Phi _{q}}(\sigma ^{\Phi })+\chi _{\Psi }(\sigma
^{\Psi }),
\end{array}
\]
where the last equality follows from the existence of a \textit{pure} state
ensemble on which the maximum in the definition of $\chi _{\Phi }(\sigma
^{\Phi })$ is achieved.$\triangle $

Due to the MSW representation \cite{MSW} and subadditivity of the quantum
entropy the subadditivity of the $\chi$-function easily follows from the
strong superadditivity of the entanglement of formation. What is interesting
that the converse is also true. The strong superadditivity of the
entanglement of formation follows from the subadditivity of the $\chi$
-function for \textit{any} two channels. Together with the corollary 3 it
provides another way (as compared to \cite{Sh-e-a-q},\cite{P}) of proving
that the global additivity conjecture for unconstrained channels (or
additivity of the entanglement of formation) implies strong superadditivity
of the entanglement of formation.

The above statement is a consequence of the following theorem.

\textbf{Theorem 2.} Let $\Phi$ and $\Psi$ be fixed channels. For
given arbitrary \break $\rho\in\mathfrak{S}(\mathcal{H})$ and
$\varrho\in\mathfrak{S}( \mathcal{K})$ the additivity conjecture
\[
\bar{C}\left( \Phi\otimes \Psi;\{\rho\}\otimes\{\varrho\}\right)=\bar{C}
(\Phi; \{\rho\})+\bar{C}(\Psi;\{\varrho\})
\]
holds if and only if
\[
\min\sum\limits_{k}\mu_{k}H(\Phi\otimes\Psi(\sigma_{k}))=\!
\min\limits_{\sum\limits_{i}\pi_{i}\rho_{i}=\rho}\sum\limits_{i}\pi_{i}H(
\Phi(\rho_{i}))+\!\!
\min\limits_{\sum\limits_{j}\varpi_{j}\varrho_{j}=\varrho}\sum\limits_{j}
\varpi_{j}H(\Psi(\varrho_{j})),
\]
where the first minimum is over all ensembles $\{\mu_{k},\sigma_{k}\}$ of
states in $\mathfrak{S}(\mathcal{H}\otimes\mathcal{K})$ such that $
\sum\limits_{k}\mu_{k}\sigma_{k}^{\Phi}=\rho$ and $\sum\limits_{k}\mu_{k}
\sigma_{k}^{\Psi}=\varrho$.

\textit{Proof.} The sufficiency of the above condition for the additivity of
the Holevo capacity for the $\{\rho \}$-constrained channel $\Phi $ and the $
\{\varrho \}$-constrained channel $\Psi $ obviously follows from the
subadditivity of quantum entropy.

Let us prove the necessity of this condition. The additivity conjecture for
the $\{\rho\}$-constrained channel $\Phi$ and $\{\varrho\}$-constrained
channel $\Psi$ implies the existence of nonentangled ensemble with average
state $\rho\otimes\varrho$, which is optimal for the $\{\rho\}\otimes\{
\varrho\}$-constrained channel $\Phi\otimes\Psi$. By proposition 2 we have
\begin{equation}  \label{opt-c}
\!\!\!\chi_{\Phi\otimes\Psi}(\rho\otimes\varrho)=
\chi_{\Phi}(\rho)+\chi_{\Psi}(\varrho)\geq\chi_{\Phi\otimes\Psi}(\sigma)+
S(\Phi\otimes\Psi(\sigma)\|\Phi(\rho)\otimes\Psi(\varrho))
\end{equation}
for any state $\sigma$ in $\mathfrak{S}(\mathcal{H})\otimes\mathfrak{S}(
\mathcal{K})$ such that $\sigma^{\Phi}=\rho$ and $\sigma^{\Psi}=\varrho$.
Note that
\begin{equation}  \label{s-c-r-e-exp}
S(\Phi\otimes\Psi(\sigma)\|\Phi(\rho)\otimes\Psi(\varrho))=
H(\Phi(\rho))+H(\Psi(\varrho))-H(\Phi\otimes\Psi(\sigma)).
\end{equation}

The inequality (\ref{opt-c}) with (\ref{s-c-r-e-exp}) and the definition of
the $\chi$-function provides $"\geq"$ in the condition of the theorem. Since
$"\leq"$ in this condition is obvious, the proof is complete.$\triangle$

\textbf{Corollary 5.} The subadditivity (\ref{sub-add}) of the function $
\chi_{\Phi\otimes\Psi}$ is equivalent to the following property:
\[
\chi_{\Phi}(\sigma^{\Phi})+\chi_{\Psi}(\sigma^{\Psi})-\chi_{\Phi\otimes
\Psi}(\sigma)\geq
H(\Phi(\sigma^{\Phi}))+H(\Psi(\sigma^{\Psi}))-H(\Phi\otimes\Psi(\sigma)).
\]
for all $\sigma\in\mathfrak{S}(\mathcal{H}\otimes\mathcal{K})$.

This means that the gap between $\chi_{\Phi}(\sigma^{\Phi})+\chi_{\Psi}(
\sigma^{\Psi})$ and $\chi_{\Phi\otimes\Psi}(\sigma)$ is no less than the gap
between $H(\Phi(\sigma^{\Phi}))+H(\Psi(\sigma^{\Psi}))$ and $
H(\Phi\otimes\Psi(\sigma))$.

\textit{Proof.} The sufficiency of the above property for inequality (\ref
{sub-add}) is clear.

To prove its necessity note that subadditivity of the function $\chi _{\Phi
\otimes \Psi }$ implies additivity for the $\{\rho \}$-constrained channel $
\Phi $ and $\{\varrho \}$-constrained channel $\Psi $ with any $\rho \in
\mathfrak{S}(\mathcal{H})$ and $\varrho \in \mathfrak{S}(\mathcal{K})$. By
the definition of the $\chi $-function the above property can be rewritten
as \newline
\newline
$\min\limits_{\sum\limits_{k}\mu _{k}\sigma _{k}=\sigma }\sum\limits_{k}\mu
_{k}H(\Phi \otimes \Psi (\sigma _{k}))\!\geq
\!\min\limits_{\sum\limits_{i}\pi _{i}\rho _{i}=\rho }\sum\limits_{i}\pi
_{i}H(\Phi (\rho _{i}))+\!\min\limits_{\sum\limits_{j}\varpi _{j}\varrho
_{j}=\varrho }\sum\limits_{j}\varpi _{j}H(\Psi (\varrho _{j})).$ with $\rho
=\sigma ^{\Phi }$ and $\varrho =\sigma ^{\Psi }$. But this inequality
follows from the theorem 2. $\triangle $

\textbf{Corollary 6.} The global subadditivity of the $\chi $-function
implies strong superadditivity of the entanglement of formation.

\textit{Proof.} Let $\;\mathcal{H}_{1},\mathcal{H}_{2},\mathcal{K}_{1},
\mathcal{K}_{2}\;$ be finite dimensional Hilbert spaces \break and $\mathcal{
\ H}=\mathcal{H}_{1}\otimes \mathcal{H}_{2}$, $\mathcal{K}=\mathcal{K}
_{1}\otimes \mathcal{K}_{2}$. Consider the channels $\Phi _{i}(\cdot )=
\mathrm{Tr}_{\mathcal{K}_{i}}(\cdot )$ from $\mathfrak{S}(\mathcal{H}
_{i}\otimes \mathcal{\ K}_{i})$ into $\mathfrak{S}(\mathcal{H}_{i})
,\;\;i=1,2 $. It is clear that $\Phi _{1}\otimes \Phi _{2}(\cdot )=\mathrm{Tr
}_{\mathcal{K}}(\cdot )$.

Let $\sigma$ be an arbitrary state in $\mathfrak{S}(\mathcal{H} \otimes
\mathcal{K})$, let $\sigma_{1}$ and $\sigma_{2}$ be the partial states of $
\sigma$, corresponding to the decomposition $\mathcal{H} \otimes\mathcal{K}
=( \mathcal{H}_{1} \otimes\mathcal{K}_{1})\otimes(\mathcal{H}_{2}\otimes
\mathcal{K}_{2})$. By definition we have
\[
\begin{array}{c}
\chi_{\Phi_{1}}(\sigma_{1})=H(\mathrm{Tr}_{\mathcal{K}_{1}}(\sigma_{1}))-
E_{F}(\sigma_{1}),\quad \chi_{\Phi_{2}}(\sigma_{2})=H(\mathrm{Tr}_{\mathcal{
K }_{2}}(\sigma_{2}))- E_{F}(\sigma_{2}), \\
\\
\chi_{\Phi_{1}\otimes\Phi_{2}}(\sigma)=H(\mathrm{Tr}_{\mathcal{K}}(\sigma))-
E_{F}(\sigma).
\end{array}
\]
By the corollary 5 with $\Phi=\Phi_{1}$ and $\Psi=\Phi_{2}$ we obtain the
desired strong superadditivity of the entanglement of formation:
\[
E_{F}(\sigma)\geq E_{F}(\sigma_{1})+ E_{F}(\sigma_{2}).\;\triangle
\]

Note also the following lower bound for the Holevo capacity of the \break $\{\rho
\}\otimes \{\varrho \}$-constrained channel $\Phi \otimes \Psi $.

\textbf{Corollary 7.} Let $\Phi $ and $\Psi $ be an arbitrary channel. For
given arbitrary states $\rho \in \mathfrak{S}(\mathcal{H})$ and $\varrho \in
\mathfrak{S}(\mathcal{K})$
\[
\bar{C}(\Phi \otimes \Psi ;\{\rho \}\otimes \{\varrho \})\!\geq \!\bar{C}
(\Phi ;\{\rho \})+\bar{C}(\Psi ;\{\varrho \})+S(\Phi (\rho )\otimes \Psi
(\varrho )\Vert \Phi \otimes \Psi (\sigma _{\mathrm{av}})),
\]
where $\sigma _{\mathrm{av}}$ is the average state of \textit{any} optimal
ensemble for the $\{\rho \}\otimes \{\varrho \}$-constrained channel $\Phi
\otimes \Psi $.

\textit{Proof.} Direct application of the proposition 2.$\triangle $

\pagebreak

\section{Application of the convex duality}

Corollary 5 of the theorem 2 and the arguments from the convex analysis
raised in \cite{A&B} provide the characterization of channels for which
subadditivity of the $\chi$-function holds. For a channel $\Phi $ and an
operator $A\in \mathfrak{B}_{+}(\mathcal{H})$ we introduce the following
modified output purity of the channel (cf. \cite{ahw})
\begin{equation}
\nu _{H}\left( \Phi ,A\right) =\min\limits_{\rho \in \mathfrak{S}(
\mathcal{H} )}\left[ H(\Phi (\rho ))+\mathrm{Tr}A\rho \right] .  \label{nu}
\end{equation}
We will show later that the additivity property for this value is equivalent
to the subadditivity of the $\chi$-function.

Let $\Phi :\mathfrak{S}(\mathcal{H})\mapsto \mathfrak{S}(\mathcal{H}^{\prime
})$ be an arbitrary channel. For a given operator $E$ in $\mathfrak{B}(
\mathcal{H}),0\leq E\leq I_{\mathcal{H}}$, and numbers $q\in \lbrack 0;1]$, $
d\in \mathbb{N}=\left\{ 1,2,\dots \right\} $ consider the channel $
\widetilde{\Phi }(E,q,d)$ from $\mathfrak{B}(\mathcal{H})$ into $\mathfrak{B}
(\mathcal{H}^{\prime })\oplus \mathbf{C}^{d+1},$ related to the Shor's
channel extension $\widehat{\Phi }(E,q,d)$ by the following equality:
\[
\widetilde{\Phi }(E,q,d)(\rho )=\widehat{\Phi }(E,q,d)(\rho \otimes \tau
),\quad \forall \rho \in \mathfrak{S}(\mathcal{H}),
\]
where $\tau $ is the chaotic classical state $[d^{-1},...,d^{-1}]$ in $C^{d}$
. We have
\[
\widetilde{\Phi }(E,q,d)(\rho )=(1-q)\Phi (\rho )\oplus q\mathbf{E}(\rho ),
\]
where $\mathbf{E}(\rho )=[\text{Tr}\rho \bar{E},d^{-1}\text{Tr}E\rho
,...,d^{-1}\text{Tr}E\rho ]$ is a channel from $\mathfrak{B}(\mathcal{H})$
into $\mathbf{C}^{d+1}$. The channel $\widetilde{\Phi }(E,q,d)$ was
originally introduced by Shor in \cite{Sh-e-a-q} with the aim to prove that
additivity of the minimum output entropy for any pair of channels implies
additivity of the entanglement of formation. We will show that subadditivity
of the function $\chi _{\Phi \otimes \Psi }$ is closely connected to
additivity of the minimum output entropy for the channels $\widetilde{\Phi }$
and $\widetilde{\Psi }$.

\textbf{Definition 2.} We say that additivity of the minimum output entropy
holds asymptotically for the sequences of channels $\{\Phi _{n}:
\mathfrak{S}( \mathcal{H})\mapsto \mathfrak{S}(\mathcal{H}_{n}^{\prime })\}$
and $\{\Psi _{m}:\mathfrak{S}(\mathcal{K})\mapsto \mathfrak{S}(
\mathcal{K}_{m}^{\prime })\}$ if
\[
\begin{array}{c}
\lim\limits_{n,m\rightarrow +\infty }\min\limits_{\sigma \in
\mathfrak{S}(\mathcal{H} \otimes \mathcal{K})}H(\Phi _{n}\otimes \Psi
_{m}(\sigma )) \\
\\
=\lim\limits_{n\rightarrow +\infty} \min\limits_{\rho\in \mathfrak{S}(
\mathcal{H} )}H(\Phi _{n}(\rho))\;+\lim\limits_{m\rightarrow +\infty
}\min\limits_{\varrho \in \mathfrak{S}(\mathcal{K})}H(\Psi _{m}(\varrho )),
\end{array}
\]
assuming that the limits exist and are finite.

Note that if additivity of the minimum output entropy holds for the channels
$\Phi _{n}$ and $\Psi _{m}$ for all sufficiently large $n$ and $m$ it
obviously holds asymptotically for the sequences $\{\Phi _{n}\}$ and $\{\Psi
_{m}\}$.

As in the case of Shor's channel extension it is convenient to denote by $
\widetilde{\Phi }_{d}(A,p)$ the channel $\widetilde{\Phi }(A,p/\log d,d)$.

\textbf{Theorem 3.} Let $\Phi :\mathfrak{S}(\mathcal{H})\mapsto
\mathfrak{S}( \mathcal{H}^{\prime })$ and $\Psi :\mathfrak{S}(\mathcal{K})
\mapsto \mathfrak{S}(\mathcal{K}^{\prime })$ be arbitrary fixed channels.
The following statements are equivalent:

\begin{enumerate}[(i)]

\item  The function $\chi _{\Phi \otimes \Psi }$ is subadditivite;

\item  For all $A\in \mathfrak{B}_{+}(\mathcal{H})$ and $B\in \mathfrak{B}
_{+}(\mathcal{K})$
\[
\nu _{H}\left( \Phi \otimes \Psi ,A\otimes I+I\otimes B\right) =\nu
_{H}\left( \Phi ,A\right) +\nu _{H}\left( \Psi ,B\right) ;
\]

\item  The additivity of the minimum output entropy holds asymptotically for
the sequences of channels $\{\widetilde{\Phi }_{d}(A,p)\}_{d\in \mathbb{N}}$
and $\{\widetilde{\Psi }_{e}(B,r)\}_{e\in \mathbb{N}}$ with arbitrary pairs $
(A,p)$ and $(B,r)$.
\end{enumerate}

\textit{Proof.} $(\textup{i})\Leftrightarrow(\textup{ii})$ Note that the
function
\[
\hat{H}_{\Phi}(\rho)=\min_{\sum\limits_{i}\pi _{i}\rho _{i}=\rho
}\sum\limits_{i}\pi _{i}H(\Phi (\rho _{i}))
\]
can be considered as the convex closure of the function $H_{\Phi}(\rho)=H(
\Phi(\rho))$, defined on the set of pure states \cite{A&B}, \cite{JT}. The
conjugate function is defined on the set $\mathfrak{B}_{h}( \mathcal{H})$ of
all hermitian operators by
\[
H^{\ast}_{\Phi}(\rho)(X)=\max_{\rho }\left[ \mathrm{Tr}X\rho
-H_{\Phi}(\rho)\right] .
\]
The inequality in the corollary 5 can be regarded as the strong
superadditivity of the function $\hat{H}_{\Phi}(\rho)$. By lemma 1 in \cite
{A&B} this superadditivity is equivalent to the subadditivity of the
conjugate function $H^{\ast}_{\Phi}$ with respect to the Kronecker sum:
\[
H^{\ast}_{\Phi}(A\otimes I_{\mathcal{K}}+I_{\mathcal{H}}\otimes B)\leq
H^{\ast}_{\Phi}(A)+H^{\ast}_{\Phi}(B),\quad \forall A\in \mathfrak{B}_{h}(
\mathcal{H}),\;\forall B\in \mathfrak{B}_{h}(\mathcal{K}).
\]
By the definition of $H^{\ast}_{\Phi}$ the last inequality is equivalent to
\[
\begin{array}{c}
\max\limits_{\sigma \in \mathfrak{S}(\mathcal{H}\otimes \mathcal{K})}\left[
\mathrm{Tr}A\sigma ^{\Phi }+\mathrm{Tr}B\sigma ^{\Psi }-H(\Phi \otimes \Psi
(\sigma ))\right] \\
\\
\leq \max\limits_{\rho \in \mathfrak{S}(\mathcal{H})}\left[ \mathrm{Tr}A\rho
-H(\Phi (\rho ))\right]\;+\max\limits_{\varrho \in \mathfrak{S}(\mathcal{K})
} \left[ \mathrm{Tr}B\varrho -H(\Psi (\varrho ))\right]
\end{array}
\]
for all $A\in \mathfrak{B}_{h}(\mathcal{H})$ and $B\in \mathfrak{B}_{h}(
\mathcal{K})$.

Noting that $"\geq"$ in the previous inequality is obvious and using the
invariance of it after changing $A$ and $B$ on $A+\|A\|I_{\mathcal{H}}$ and $
B+\|B\|I_{\mathcal{K}}$ correspondingly we obtain that $(\textup{i}
)\Leftrightarrow(\textup{ii})$.

$(\textup{ii})\Leftrightarrow(\textup{iii})\;$ It is necessary to obtain
the expressions for $\,H(\widetilde{ \Phi }_{d}(A,p)(\rho ))\,$, $\,H(
\widetilde{\Psi }_{e}(B,r)(\varrho ))\,$ and $\,H( \widetilde{\Phi }
_{d}(A,p)\otimes \widetilde{\Psi }_{e}(B,r)(\sigma ))\,$ for arbitrary $\rho
\in \mathfrak{S}(\mathcal{H})$, $\varrho \in \mathfrak{S}( \mathcal{K})$ and
$\sigma \in \mathfrak{S}(\mathcal{H}\otimes \mathcal{K})$. For this aim we
will use the following simple lemma.

\textbf{Lemma 2.} Let $\{\Phi _{j}\}_{j=1}^{n}$ be a collection of channels
from $\mathfrak{S}(\mathcal{H})$ into $\mathfrak{S}(\mathcal{H}_{j}^{\prime
})$, $\{q_{j}\}_{j=1}^{n}$ be a probability distribution. Then for the
channel $\Phi =\bigoplus_{j=1}^{n}q_{j}\Phi _{j}$ from $\mathfrak{S}(
\mathcal{H})$ into $\mathfrak{S}(\bigoplus_{j=1}^{n}\mathcal{H}_{j}^{\prime
})$ one has
\[
H(\Phi (\omega ))=H(\{q_{j}\})+\sum_{j=1}^{n}q_{j}H(\Phi _{j}(\omega
)),\quad \forall \omega \in \mathfrak{S}(\mathcal{H}).
\]

Applying this lemma to the channels $\widetilde{\Phi }_{d}(A,p)=(1-q^{\prime
})\Phi +q^{\prime }\mathbf{A}$ and $\widetilde{\Psi }_{e}(B,r)=(1-q^{\prime
\prime })\Psi +q^{\prime \prime }\mathbf{B}$ (where $q^{\prime }=p/\log d$
and $q^{\prime \prime }=r/\log e$) we obtain
\begin{eqnarray*}
H(\widetilde{\Phi }_{d}(A,p)(\rho )) &=&h_{2}(q^{\prime })+(1-q^{\prime
})H(\Phi (\rho ))+p\mathrm{Tr}A\rho +q^{\prime }h_{2}(\mathrm{Tr}A\rho ), \\
H(\widetilde{\Psi }_{e}(B,r)(\varrho )) &=&h_{2}(q^{\prime \prime
})+(1-q^{\prime \prime })H(\Psi (\varrho ))+r\mathrm{Tr}B\varrho +q^{\prime
\prime }h_{2}(\mathrm{Tr}B\varrho ).
\end{eqnarray*}
The above expressions imply existence of the limits:
\begin{eqnarray}
\lim_{d\rightarrow +\infty }\min_{\rho \in \mathfrak{S}(\mathcal{H})}H(
\widetilde{\Phi }_{d}(A,p)(\rho )) &=&\min_{\rho \in \mathfrak{S}(
\mathcal{H} )}\left[ H(\Phi (\rho ))+p\mathrm{Tr}A\rho \right] ,
\label{lim-exp-1} \\
\lim_{e\rightarrow +\infty }\min_{\varrho \in \mathfrak{S}(\mathcal{K})}H(
\widetilde{\Psi }_{e}(B,r)(\varrho )) &=&\min_{\varrho \in \mathfrak{S}(
\mathcal{K})}\left[ H(\Psi (\varrho ))+r\mathrm{Tr}B\varrho \right] .
\label{lim-exp-2}
\end{eqnarray}
Let us prove that
\begin{equation}
\begin{array}{c}
\lim\limits_{d,e\rightarrow +\infty }\min\limits_{\sigma \in \mathfrak{S}(
\mathcal{H}\otimes\mathcal{K})}H(\widetilde{\Phi }_{d}(A,p)\otimes
\widetilde{\Psi }_{e}(B,r)(\sigma )) \\
\\
=\min\limits_{\sigma \in \mathfrak{S}(\mathcal{H}\otimes\mathcal{K})}\left[
H(\Phi \otimes \Psi (\sigma ))+\mathrm{Tr}A\sigma ^{\Phi }+\mathrm{Tr}
B\sigma ^{\Psi }\right] .
\end{array}
\label{lim-exp-3}
\end{equation}
Due to the representation
\begin{equation}
\begin{array}{c}
\widetilde{\Phi }_{d}(A,p)\otimes \widetilde{\Psi }_{e}(B,r)=(1-q^{\prime
})(1-q^{\prime \prime })\Phi \otimes \Psi  \\
\\
+q^{\prime }(1-q^{\prime \prime })\mathbf{A}\otimes \Psi +(1-q^{\prime
})q^{\prime \prime }\Phi \otimes \mathbf{B}+q^{\prime }q^{\prime \prime }
\mathbf{A}\otimes \mathbf{B}
\end{array}
\label{d-s-r}
\end{equation}
the lemma 2 reduces the calculation of the value $H(\widetilde{\Phi }
_{d}(A,p)\otimes \widetilde{\Psi }_{e}(B,r)(\sigma ))$ to the calculation of
four entropies, the first of which is $H(\Phi \otimes \Psi (\sigma )),$
while the fourth does not exceed $\log (d+1)+\log (e+1)$. To calculate the
others note that
\[
\mathbf{A}\otimes \Psi (\sigma )=[\Psi _{\bar{A}}(\sigma ),d^{-1}\Psi
_{A}(\sigma ),...,d^{-1}\Psi _{A}(\sigma )],
\]
where $\Psi _{X}(\cdot )=\mathrm{Tr}_{\mathcal{H}}\left( X\otimes I_{
\mathcal{K}}\cdot (\textup{Id}\otimes \Psi )(\cdot )\right) $ is a
completely positive map from $\mathfrak{S}(\mathcal{H\otimes K})$ into $
\mathfrak{B}_{+}(\mathcal{K^{\prime }})$ ($X=A,\bar{A}$ and $\textup{Id}$
is an identity map on $\mathfrak{S}(\mathcal{H})$). Therefore,
\begin{equation}
\begin{array}{c}
H(\mathbf{A}\otimes \Psi (\sigma ))=H(\Psi _{\bar{A}}(\sigma ))+H(\Psi
_{A}(\sigma ))+\log d\mathrm{Tr}\Psi _{A}(\sigma ) \\
\\
=F(\sigma |A,\Psi )+\log d\mathrm{Tr}A\sigma ^{\Phi },\label{A}
\end{array}
\end{equation}
where $F(\sigma |A,\Psi )$ does not depend on $d$ and does not exceed $2\log
\dim \mathcal{K}^{\prime }$.\newline
Similarly,
\begin{equation}
H(\Phi \otimes \mathbf{B}(\sigma ))=G(\sigma |B,\Phi )+\log e\mathrm{Tr}
B\sigma ^{\Psi },  \label{B}
\end{equation}
where $G(\sigma |B,\Phi )$ does not depend on $e$ and does not exceed $2\log
\dim \mathcal{H}^{\prime }$.

The lemma 2 and the representation (\ref{d-s-r}) imply
\[
\begin{array}{c}
H(\widetilde{\Phi }_{d}(A,p)\otimes \widetilde{\Psi }_{e}(B,r)(\sigma
))=H(\{q^{\prime }q^{\prime \prime },q^{\prime }(1\!-\!q^{\prime \prime
}),q^{\prime \prime }(1\!-\!q^{\prime }),(1\!-\!q^{\prime })(1\!-\!q^{\prime
\prime })\}) \\
\\
+(1-q^{\prime })(1-q^{\prime \prime })H(\Phi \otimes \Psi (\sigma
))+(1-q^{\prime \prime })p\mathrm{Tr}A\sigma ^{\Phi }+(1-q^{\prime })r
\mathrm{Tr}B\sigma ^{\Psi } \\
\\
+(1-q^{\prime \prime })q^{\prime }F(\sigma |A,\Psi )+(1-q^{\prime
})q^{\prime \prime }G(\sigma |B,\Phi )+q^{\prime }q^{\prime \prime }H(
\mathbf{A}\otimes \mathbf{B}(\sigma )).
\end{array}
\]
The estimates for $F,G$ and
\[
q^{\prime }q^{\prime \prime }H(\mathbf{A}\otimes \mathbf{B}(\sigma ))\leq
\frac{pr(\log (d+1)+\log (e+1))}{\log d\log e}.
\]
imply that the right side of the above expression tends to
\[
H(\Phi \otimes \Psi (\sigma ))+\mathrm{Tr}A\sigma ^{\Phi }+\mathrm{Tr}
B\sigma ^{\Psi }
\]
uniformly in $\sigma $ when $d$ and $e$ tend to infinity. This proves (\ref
{lim-exp-3}).

The expressions (\ref{lim-exp-1}),(\ref{lim-exp-2}) and (\ref{lim-exp-3})
show that the statement $(\textup{iii})$ is equivalent to the equality
\[
\begin{array}{c}
\min\limits_{\sigma \in \mathfrak{S}(\mathcal{H}\otimes \mathcal{K})}\left[
H(\Phi \otimes \Psi (\sigma ))+p\mathrm{Tr}A\sigma ^{\Phi }+r\mathrm{Tr}
B\sigma ^{\Psi }\right] \\
\\
=\min\limits_{\rho \in \mathfrak{S}(\mathcal{H})}\left[ H(\Phi (\rho ))+p
\mathrm{Tr}A\rho \right] +\min\limits_{\varrho \in \mathfrak{S}(\mathcal{K})
} \left[ H(\Psi (\varrho ))+r\mathrm{Tr}B\varrho \right]
\end{array}
\]
for arbitrary nonnegative numbers $p,r$ and operators $A,B$ such
that \break $0\leq A\leq I_{\mathcal{H}},\;0\leq B\leq
I_{\mathcal{K}}$. But this is exactly the statement
$(\textup{ii})$. $\triangle$

The concavity argument shows that the minimum in (\ref{nu}) is achieved on a
pure state. So, the statement $(\textup{ii})$ of the above theorem
characterizes the subadditivity of the $\chi$ -function for the particular
channels in terms of action of these channels and its tensor product on pure
states.

Note also that $(\textup{ii})$ implies additivity of the minimum output
entropy (the case $A=B=0$).

\textbf{Corollary 8.} Additivity of the minimum output entropy for the
channels $\widetilde{\Phi }_{d}(A,p)$ and $\widetilde{\Psi }_{e}(B,r)$ with
arbitrary pairs $(A,p)$ and $(B,r)$ for all sufficiently large $d$ and $e$
implies subadditivity of the function $\chi _{\Phi \otimes \Psi }$.

\textbf{Corollary 9.} Global additivity of the minimum output entropy is
equivalent to the global subadditivity of the $\chi $ -function.

\textit{Proof.} By the corollary in previous section, subadditivity of the
function $\chi _{\Phi \otimes \Psi }$ implies additivity of the minimum
output entropy for the channels $\Phi $ and $\Psi $.

If additivity of the minimum output entropy holds for any pair of channels
then, by corollary 8, the $\chi $-function is globally subadditive.$
\triangle $

\textbf{Corollary 10.} The asymptotic additivity of the minimum output
entropy for the sequences of channels $\{\widetilde{\Phi }_{d}(A,p)\}_{d\in
\mathbb{N}}$ and $\{\widetilde{\Psi }_{e}(B,r)\}_{e\in \mathbb{N}}$ with
arbitrary pairs $(A,p)$ and $(B,r)$ is equivalent to the asymptotic
additivity of the Holevo capacity for the sequences of channels $\{\widehat{
\Phi }_{d}(A,p)\}_{d\in \mathbb{N}}$ and $\{\widehat{\Psi }_{e}(B,r)\}_{e\in
\mathbb{N}}$ with arbitrary pairs $(A,p)$ and $(B,r)$.

\textit{Proof.} By theorems 1 and 3 both properties are equivalent to
subadditivity of the function $\chi _{\Phi \otimes \Psi }$. $\triangle $

\textbf{Remark.} The proof of theorem 3 can be modified to obtain a direct
proof of the fact that additivity of the minimum output entropy for any pair
of channels implies the strong subadditivity of the entanglement of
formation (instead of simple additivity as in \cite{Sh-e-a-q}). To see this
consider channels $\Phi $ and $\Psi $ of the form of the partial traces, as
in the proof of the corollary 6. The arguments in the proof of theorem 3
shows that assumed additivity of the minimum output entropy for the channels
$\widetilde{\Phi }_{d}(A,p)$ and $\widetilde{\Psi }_{e}(B,r)$ implies
superadditivity of the function $H_{\Phi \otimes \Psi }(\sigma )$, which in
this case coincides with the entanglement of formation $E_{F}(\sigma )$.

\section{Some relations for the relative entropy}

The proofs of the theorem 2 and corollary 7 were based on proposition 2. In
this section we provide direct proofs of them considering some interesting
relations for the relative entropy.

Let $\sigma$ and $\varsigma$ be arbitrary states such that $\mathrm{supp}
\sigma\subseteq\mathrm{supp}\varsigma$. The functions $f(x)=S(x\sigma+(1-x)
\varsigma\|\varsigma)$ and $g(x)=S(\varsigma\|x\sigma+(1-x)\varsigma)$ are
obviously continuous and convex on $[0;1\}\footnote{
The symbol $\}$ means $)$ if $\mathrm{supp}\sigma\subset\mathrm{supp}
\varsigma$ and $]$ if $\mathrm{supp}\varsigma=\mathrm{supp}\sigma$}$.

\textbf{Proposition 5.} These functions are related by the following
transformations:
\[
g(x)=xf^{\prime}(x)-f(x),\quad f(x)=x\int\limits_{0}^{x}\frac{g(t)}{t^{2}}
dt,\quad\quad \forall x\in[0;1\}
\]
with the "initial conditions" $f(0)=g(0)=0$ and
\begin{equation}  \label{der-in-0}
f^{\prime}(0)=\frac{d}{dx}S(x\sigma+(1-x)\varsigma\|\varsigma)|_{x=0}=0.
\end{equation}

The proof of this proposition and some features of the above transformations
can be found in the Appendix.

Note that in the case $\mathrm{supp}\sigma=\mathrm{supp}\varsigma$
proposition 5 implies
\begin{equation}  \label{der-in-1}
f^{\prime}(1)=\frac{d}{dx}S(x\sigma+(1-x)\varsigma\|\varsigma)|_{x=1}=
S(\sigma\|\varsigma)+S(\varsigma\|\sigma).
\end{equation}

We will use proposition 5 with $\Phi \otimes \Psi (\sigma )$ and $\Phi
(\sigma ^{\Phi })\otimes \Psi (\sigma ^{\Psi })$ in the role of $\sigma $
and $\varsigma $ correspondingly. Taking into account the definition of the
partial trace one can verify that $\mathrm{supp}\,\Phi \otimes \Psi (\sigma
)\subseteq \mathrm{supp}\,\Phi (\sigma ^{\Phi })\otimes \Psi (\sigma ^{\Psi
})\;(\forall \sigma )$. In this case
\[
f(x)=H(\Phi (\sigma ^{\Phi }))+H(\Psi (\sigma ^{\Psi }))-H(x\Phi \otimes
\Psi (\sigma )+(1-x)\Phi (\sigma ^{\Phi })\otimes \Psi (\sigma ^{\Psi })).
\]
The equality (\ref{der-in-0}) may be used for proving theorem 2 while (\ref
{der-in-1}) provides the statement of the corollary 7.

Let $\{\pi _{i},\rho _{i}\}$ and $\{\varpi _{j},\varrho _{j}\}$ be the
optimal ensembles for the $\{\rho \}$-constrained channel $\Phi $ and the $
\{\varrho \}$-constrained channel $\Psi$ (with the average states $\rho$ and
$\varrho$) correspondingly. Consider the mixture of an arbitrary ensemble $
\{\mu _{k},\sigma _{k}\}$ (with the average state $\sigma _{\mathrm{av}}$)
and the ensemble $\,\{\pi _{i}\varpi _{j},\rho _{i}\otimes \varrho _{j}\}\,$
with the weights $x\,$ and $1-x$ correspondingly. This new ensemble has the
average state \break $x\sigma _{\mathrm{av}}+(1-x)\rho \otimes \varrho \in
\{\rho \}\otimes \{\varrho \}$ and the Holevo quantity
\[
\begin{array}{c}
\chi _{\Phi \otimes \Psi }^{x}=H(\Phi \otimes \Psi (x\sigma _{\mathrm{av}
}+(1-x)\rho \otimes \varrho ))-x\sum\limits_{k}\mu _{k}H(\Phi \otimes \Psi
(\sigma _{k})) \\
-(1-x)\left( \sum\limits_{i}\pi _{i}H(\Phi (\rho
_{i}))+\sum\limits_{j}\varpi _{j}H(\Psi (\varrho _{j})\right) \\
\\
=\chi _{\Phi }(\rho )+\chi _{\Psi }(\varrho )+H(\Phi \otimes \Psi (x\sigma _{
\mathrm{av}}+(1-x)\rho \otimes \varrho ))-H(\Phi (\rho ))-H(\Psi (\varrho ))
\\
\\
+x\left( \sum\limits_{i}\pi _{i}H(\Phi (\rho _{i}))+\sum\limits_{j}\varpi
_{j}H(\Psi (\varrho _{j})-\sum\limits_{k}\mu _{k}H(\Phi \otimes \Psi (\sigma
_{k}))\right)
\end{array}
\]
Denoting the quantity in the last brackets by $\Delta $ we obtain
\begin{equation}
h(x)=\chi _{\Phi \otimes \Psi }^{x}-\chi _{\Phi \otimes \Psi }^{0}=x\Delta
-f(x),  \label{delta-chi}
\end{equation}
where $f(x)=H(\Phi (\rho ))+H(\Psi (\varrho ))-H(\Phi \otimes \Psi (x\sigma
_{\mathrm{av}}+(1-x)\rho \otimes \varrho ))$ is concave function with $
f^{\prime }(0)=0$ by proposition 5. If the value $\Delta $ is positive,
then, due to $f^{\prime }(0)=0$, we will necessarily have $\chi _{\Phi
\otimes \Psi }^{x}>\chi _{\Phi \otimes \Psi }^{0}$ for sufficiently small $x$
. If the value $\Delta $ is not positive, then $\chi _{\Phi \otimes \Psi
}^{x}<\chi _{\Phi \otimes \Psi }^{0}$ for all $x>0$. This observation proves
the theorem 2.

Consider the above construction in the case where $\{\mu_{k},\sigma_{k}\}$
is an optimal ensemble for the $\{\rho\}\otimes\{\varrho\}$-constrained
channel $\Phi\otimes\Psi$. In this case proposition 1 implies $\textup{supp}
\,\Phi\otimes\Psi(\sigma_{\mathrm{av}})= \textup{supp}\,\Phi(\rho)\otimes
\Phi(\varrho)$. Hence, by proposition 5, the function $f(x)$ is defined on $
[0, 1]$ and
\begin{equation}  \label{der-exp}
f^{\prime}(1)=S(\Phi\otimes\Psi(\sigma_{\mathrm{av}})\|
\Phi(\rho)\otimes\Psi(\varrho)) +S(\Phi(\rho)\otimes\Psi(\varrho)\|
\Phi\otimes\Psi(\sigma_{\mathrm{av}})).
\end{equation}
Due to optimality of the above ensemble the concave function $h(x)$, defined
by (\ref{delta-chi}), must be nondecreasing on $[0, 1]$ and hence $
h^{\prime}(x)=\Delta-f^{\prime}(x)\geq 0$. By concavity, this implies $
f^{\prime}(1)\leq\Delta$. But in this case
\[
\begin{array}{c}
\Delta=\sum\limits_{i}\pi_{i}H(\Phi(\rho_{i}))+
\sum\limits_{j}\varpi_{j}H(\Psi(\varrho_{j})-\sum\limits_{k}\mu_{k}H(\Phi
\otimes\Psi(\sigma_{k})) \\
\\
= \chi_{\Phi\otimes\Psi}(\sigma)-\chi_{\Phi}(\rho)-\chi_{\Psi}(\varrho)+
S(\Phi\otimes\Psi(\sigma_{\mathrm{av}})\| \Phi(\rho)\otimes\Psi(\varrho)) \\
\\
=\bar{C}( \Phi\otimes \Psi;\{\rho\}\otimes\{\varrho\})-\bar{C}(\Phi;
\{\rho\})-\bar{C}(\Psi;\{\varrho\}) \\
\\
+ S(\Phi\otimes\Psi(\sigma_{\mathrm{av}})\| \Phi(\rho)\otimes\Psi(\varrho)).
\end{array}
\]
Taking into account (\ref{der-exp}) we obtain that the inequality $
f^{\prime}(1)\leq\Delta$ is equivalent to the inequality in the corollary 7.

\section{Appendix}

For proving the proposition 5 it is sufficient to show that $f^{\prime}(0)=0$
and $f^{\prime}(x)=\langle f(x)+g(x)\rangle/x$ for $x>0$. By definition
\begin{equation}  \label{der}
\begin{array}{c}
f^{\prime}(x_{0})=\frac{d}{dx}\left[\mathrm{Tr}(x\sigma+(1-x)\varsigma)
\log(x\sigma+(1-x)\varsigma)\right]|_{x=x_{0}} \\
\\
- \frac{d}{dx}\left[\mathrm{Tr}(x\sigma+(1-x)\varsigma) \log\varsigma\right]
|_{x=x_{0}}= \mathrm{Tr}(\sigma-\varsigma)\log(x\sigma+(1-x)\varsigma) \\
\\
+ \frac{d}{dx}\left[\mathrm{Tr}(x_{0}\sigma+(1-x_{0})\varsigma)
\log(x\sigma+(1-x)\varsigma)\right]|_{x=x_{0}}- \mathrm{Tr}
(\sigma-\varsigma)\log\varsigma \\
\\
= \mathrm{Tr}(\sigma-\varsigma)(\log(x\sigma+(1-x)\varsigma)-\log\varsigma),
\end{array}
\end{equation}
where we use $\frac{d}{dx}\left[\mathrm{Tr}(x_{0}\sigma+(1-x_{0})\varsigma)
\log(x\sigma+(1-x)\varsigma)\right]|_{x=x_{0}}\!=\!0$ due to the fact that $
x_{0}$ is the maximum point of the function
\[
x\mapsto \mathrm{Tr}(x_{0}\sigma+(1-x_{0})\varsigma)\log(x\sigma+(1-x)
\varsigma).
\]
Expression (\ref{der}) with $x_{0}=0$ gives $f^{\prime}(0)=0$. By the
definition
\[
f(x)+g(x)=x\mathrm{Tr}(\sigma-\varsigma)(\log(x\sigma+(1-x)\varsigma)-\log
\varsigma).
\]
Comparing this with (\ref{der}) completes the proof.$\triangle$

The transformation $f(x)\mapsto g(x)=xf^{\prime}(x)-f(x)$ has simple
geometric meaning and can be represented as $g(x)=f^{*}(f^{\prime}(x))$,
where $f^{*}$ is the Legendre transform of $f$ \cite{JT}. The eigenvectors
of this transformation are powers $x^{\alpha}$ with corresponding
eigenvalues $1-\alpha$.

The inequality for relative entropy $S(\sigma^{\prime}\|\sigma^{\prime
\prime})\geq\frac{1}{2}\|\sigma^{\prime}-\sigma^{\prime\prime}\|^{2}\;$ \cite
{O&P} shows that
\[
f(x)\geq cx^{2},\quad g(x)\geq cx^{2},\quad c=\frac{1}{2}\|\sigma-\varsigma
\|^{2}.
\]
It is interesting to note that the "bound" $cx^{2}$ is (essentially unique)
fixed point of the above transformations.

Proposition 5 implies that if $\mathrm{supp}\sigma \subset \mathrm{supp}
\varsigma $ then
\[
g(x)\rightarrow +\infty \quad \textup{as}\quad x\rightarrow 1,\quad \mathrm{
\ but}\quad \int\limits_{0}^{1}g(x)dx<+\infty .
\]

Note also the following relations for the derivatives
\[
\frac{d^{n}}{dx^{n}}g(x)|_{x=0}=(n-1)\frac{d^{n}}{dx^{n}}f(x)|_{x=0},\quad
n\in \mathbb{N}.
\]
It implies that $f(x)-g(x)=o(x^{2})$ in a neighborhood of zero and its sign
is defined by the sign of the third derivative at zero of $f(x)$. \vspace{
15pt}

\textbf{Acknowledgments}

The author is grateful to A.S.Holevo for stimulating this work and permanent
help. This work was partially supported by INTAS grant 00-738.

\end{document}